\documentclass[fleqn,usenatbib]{mnras}

\usepackage{newtxtext,newtxmath}

\usepackage[T1]{fontenc}

\DeclareRobustCommand{\VAN}[3]{#2}
\let\VANthebibliography\thebibliography
\def\thebibliography{\DeclareRobustCommand{\VAN}[3]{##3}\VANthebibliography}

\usepackage{graphicx}	
\usepackage{amsmath}	
\usepackage{subcaption}
\usepackage[xcolor={divpdf,grey},authormarkup=none]{changes} 
\usepackage{comment}
\usepackage{float}

\newcommand{\detectifz}{\mbox{{\sc \small DETECTIFz}}}

\newcommand{\stkout}[1]{\ifmmode\text{\sout{\ensuremath{#1}}}\else\sout{#1}\fi}
\setdeletedmarkup{\stkout{#1}}
\definechangesauthor[name={Florian Sarron},color=orange]{FS}
\definechangesauthor[name={Chris Conselice},color=red]{CC}

\title[Galaxy groups at $z < 2.5$ in the REFINE survey]{DETECTIFz galaxy groups in the REFINE survey - 1. Group detection and quenched fraction evolution at $z < 2.5$}

\author[F. Sarron et al.]{
Florian Sarron,$^{1,2}$\thanks{E-mail: florian.sarron@manchester.ac.uk (FS)}
Chrisopher J. Conselice,$^{1,2}$\thanks{E-mail:conselice@manchester.ac.uk (CJC)}
\\
$^{1}$School of Physics and Astronomy, University of Nottingham, Nottingham NG7 2RD, UK\\
$^{2}$ Jodrell Bank Centre for Astrophysics, University of Manchester, Oxford Road, Manchester UK
}

\date{Accepted  2021 June 19. Received  2021 June 18; in original form 2021 May 2}

\pubyear{2021}

\begin{document}
\label{firstpage}
\pagerange{\pageref{firstpage}--\pageref{lastpage}}
\maketitle

\begin{abstract}
We use a large K-selected sample of 299,961 galaxies from the REFINE survey, consisting of a combination of data from three of the deepest near-infrared surveys: UKIDSS UDS, COSMOS/UltraVISTA and CFHTLS-D1/VIDEO, that were homogeneously reduced to obtain photometric redshifts and stellar masses. We detect 2588 candidate galaxy groups up to $z=3.15$ at $S/N>1.5$. We build a very pure ($>90\%$) sample of 448 candidate groups up to $z=2.5$ and study some of their properties. Cluster detection is done using the DElaunay TEssellation ClusTer IdentiFication with photo-z (\detectifz) algorithm that we describe. This new group finder algorithm uses the joint probability distribution functions (PDF) of redshift and stellar-mass of galaxies to detect groups as stellar-mass overdensities in overlapping redshift slices, where density is traced using Monte Carlo realisation of the Delaunay Tessellation Field Estimator (DTFE). We compute the algorithm selection function using mock galaxy catalogues taken from cosmological N-body simulation lightcones. Based on these simulations, we reach a completeness of $\sim80\%$ for clusters ($M_{200}>10^{14}{\rm M}_{\sun}$) at a purity of $\sim90\%$ at $z<2.5$. Using our 403 most massive candidate groups, we constrain the redshift evolution of the group galaxy quenched fraction at $0.12\le z<2.32$, for galaxies with $10.25 < \log M_\star/{\rm M}_{\sun} < 11$ in $0.5\times R_{200}$. We find that the quenched fraction in group cores is higher than in the field in the full redshift range considered, the difference growing with decreasing redshift. This indicates either more efficient quenching mechanisms in group cores at lower redshift or pre-processing by cosmic filaments.
\end{abstract}




\section{Introduction}

The formation and evolution of galaxies, and the clusters and groups where a large fraction of them inhabit and are intimately related, have implications for many different areas of astrophysics and cosmology. Ever since the earliest studies by Messier (1781), it has increasingly becoming established that galaxies cluster and group together and are not just randomly distributed on the sky. Soon after this there were hints of an evolutionary connection between galaxies and their environments. The major work which set off the modern study of this relation was \citet{Dressler1980} who showed that galaxies are more likely to be passive, older, and early-type in dense local environments than in lower density ones. This later has been expanded to include other features of galaxies such as star formation \citet{Gomez2003}, such that it is clear that galaxies in dense environments have a different star formation rate than similar mass galaxies in lower density environments.\\
\indent Therefore for decades it has been clear that galaxies are more evolved in denser areas than in lower density environments, at least at redshifts $z < 1$.  This implies that star formation is quenched earlier, or somehow does not continue, in galaxies found in high density environments compared with those in low density environments. In general either galaxies finish their star formation earlier and no gas is replenished in dense environments due to gas exhaustion/strangulation \citep[][]{Larson1980}, or the environment itself is driving the reduction or removal of star formation such as through process including ram-pressure stripping \citep{Gunn1972} and high-speed galaxy encounters \citep{Moore1996}. Many recent studies find this relation appears to hold even at $1 < z < 1.5$, some of the highest redshifts in which large enough samples of clusters can be found \citep[see e.g.][]{Quadri2012,Cooke2016,Nantais2016,Papovich2018,VdB2020}. However, there is considerable debate about the existence of a turn-over in this relationship at higher redshifts where clusters are not so easily found \citep[e.g.][]{Elbaz2007,Lani2013,Nantais2016}, and thus the picture is not as clear for lower density environment such as groups.\\
\indent Major questions relating to the formation of the earliest clusters pertain not only to the formation and quenching of galaxies, but also to the existence and formation of clusters themselves which can have cosmological implications \citep[see][for a review]{Allen+11}. In general, we find that galaxy evolution occurs via both internal and external mechanisms and forces.  The environments in which a galaxy finds itself must have a strong effect on how it forms and evolves, simply due to the range of environments galaxies are located within and the physical effects of those environments. An example of this are galaxies surrounded by other galaxies in close proximity such as those involved in galaxy mergers \citep[e.g.][]{Duncan2019}.  In these cases gravitational interactions and mergers induce star formation and dynamical processes that can, in the right conditions, remove stars/gas through tides. As such, the masses of galaxies can increase through accretion of satellites and the number of  galaxies decreases due to mergers. Furthermore, there can be changes to the structures and morphologies of galaxies due to this close proximity and high-speed encounters \citep[e.g.][]{Mastropietro2005}. Dense areas such as clusters also often contain an intracluster medium -- gas in the space between the galaxies in a group. This intracluster gas can interact with the gas located within galaxies themselves through ram-pressure stripping. This mechanism strips the galaxy of its cold gas, thus preventing further star-formation. This process, that has been extensively observed in galaxy clusters \citep[e.g.][]{Scott2012,Gavazzi2018,Vulcani2020} can be very efficient in this environment, shutting down star formation on time-scales of tens of Myr \citep[][]{Abadi1999}, and it may be the dominant galaxy evolution mechanism in massive clusters \citep[see][for a review]{Boselli2014}. Even though this mechanism has been recently observed in galaxy groups in the local universe \citep[][]{Vulcani2018}, its efficiency in groups is thought to be limited \citep[e.g.][]{Rasmussen2006}. However, the lower density intragroup gas can still efficiently strip the gaseous halo surrounding galaxies that is not as strongly gravitationally bound to it. When this happens, and because gas cannot be replenished after all the existing gas is used up in star formation, a galaxy will become 'passive' and lose its structure due to faded star formation \citep[e.g.][]{Wetzel2013,Peng2015}. Even lower density cosmic web filaments have been shown to be favourable environments for suppressing star-formation \citep[e.g.][]{Kuutma2017,Laigle2018,Kraljic2018} with different quenching mechanisms proposed \citep{AragonCalvo2019,Song2021}.\\

\indent Previous results show that for the most part the properties of galaxies are determined by both their environment and the individual mass of a galaxy \citep[e.g.][at $z < 1$]{Peng+10}.  Therefore, there are two ways to prevent further star formation from taking place within a galaxy - either a dense environment, which due to a rich intracluster environment quenches the star formation or quenching due to containing a high mass - so called 'mass quenching' \citep[e.g.][]{Peng+10,Bluck2019}.  For the lowest mass galaxies it is likely that the rich environment is mostly responsible for quenching, but for higher mass galaxies, the situation is more complicated \citep[e.g.][]{Grutzbauch2011}.  Moreover, while the efficiency of massive galaxy clusters in quenching star-formation at $z > 1$ has become clearer in recent years \citep[][]{Nantais2016,VdB2020}, the ability of high redshift groups to do so is still unclear. Similarly, examining the morphology-density relationship at $z < 1$, \citet{Tasca2009} showed that the morphology-density relation evolves slightly with redshift, becoming flatter and less strong at higher-z \citep[see also][]{Grutzbauch2011, Grutzbauch2011b}, and is most likely responsible for producing low mass galaxy morphology, while stellar-mass seems to play a more important role for massive galaxy morphology. The morphology results are also backed up by finding a kinematic-structure/environment relation \citet{Brough2017}. The question however remains,  which relationship - mass or environment - is more fundamental \citep[e.g.][]{Grutzbauch2011}? This relates to, and is another way of addressing, the nature vs. nurture problem for galaxy formation. \\

\indent We can address this question by examining the stellar populations, structures, and morphologies of galaxies within higher redshift overdensities, or  within proto-clusters that are just forming \citep[e.g.][]{Dressler1997,Holden2007,Sazonova2020}. However, observations of overdensities at high redshifts $z > 1.5$ are just starting in earnest, but already the observations available gives us some ideas of how this evolves. What we know so far is that massive clusters  at high redshifts,  up to $z \sim 1.2$, display a similar morphology-density relation as local galaxies, such that  denser areas of galaxies contain higher fractions of elliptical systems.  However, what is also seen is that the normalisation is lower, such that at a given density there are fewer ellipticals than at the same local density at lower redshifts \citep[e.g.][]{Sazonova2020}. It is also known that the quenched fraction is higher in massive clusters up to $z = 1.5$ \citep[e.g.][]{Lin2017,Sarron2018,VdB2018,VdB2020}. This is seen as a stronger effect for low mass galaxies, but steadily decreases with  increasing redshift. At higher redshifts $1.5 < z < 2$, few massive clusters have been studied but these can exhibit quenched fractions as high as $\sim 75\%$ \citep{Newman2014,Strazzullo2019}\\

\indent What is now needed is a systematic search for distant clusters and groups, both as a way to find clusters and to study their galaxy populations, but also to determine the best ways to maximise the purity and completeness of samples of groups/clusters. There have been few such studies \citep[e.g.][]{Papovich2008,Galametz2012,Chiang2014,Rettura2014,Ando2020} and many of them are biased by severe selection effects due to detection process targeting specific galaxy properties \citep[e.g. colour selection or radio activity, see][for a review on high redshift overdensity and (proto-)cluster detection]{Overzier2016}. Among the many ways to search for distant galaxy clusters, perhaps the cleanest way is to search for overdensities of galaxies in a limited area. In this sense we are simply looking for multiple galaxies that have similar redshifts and are found in the same part of the sky.\\

\indent Thus, in this paper we carry out a new analysis of the three deepest extragalactic deep fields measured to date - the UltraVISTA, the UKIDSS Ultra-Deep Survey (UDS) and the VIDEO surveys to find the most distant and massive clusters within the deepest ground based imaging fields. In many ways, this analysis is a precursor to what can be done with large forthcoming imaging surveys, such as Euclid, Rubin and Roman. Our goals in this paper are to provide a catalogue of sources, analyse the likelihood of galaxies being members of clusters through simulations, and provide a frame-work for investigating the ability to find $z > 1.5$ groups and clusters and study galaxy properties therein in future deep surveys such as Euclid and Rubin which will have similar depths, but cover thousands of times more area.\\

\indent This paper is organized as follows: In Sect.~\ref{sec:data} we explain the data that we use in this paper, including the mock data. In Sect.~\ref{sec:DETECTIFz} we explain our new algorithm - the \detectifz~method for finding overdensities. Sect.~\ref{DETECTIFz_cat} includes our group and cluster finding results, including a discussion of the catalogue of sources which we find. Sect.~\ref{sec:fQ} presents the evolution of the quenched fraction for galaxies as a function of redshift which are discussed in Sect.~\ref{sec:discussion}. We use AB magnitudes throughout the paper and assume a flat $\Lambda$CDM cosmology with $\Omega_{\rm m} = 0.3$ and $h = 0.7$.

\section{Data and mock data}\label{sec:data}

The data products for this study are those which arise from the REFINE (Redshift Evolution and Formation In Extragalactic systems) project, which is essentially a re-derivation and homogenisation of the major ground-based data sets used to study the distant universe \citep[see e.g.][]{Mundy2017}. As such, this paper is based on the data products presented in \citet{Mundy2017} for the UKIDSS Ultra Deep Survey (UDS), COSMOS/UltraVISTA and CHFTLS-D1/VIDEO survey regions. \citet{Mundy2017} computed photometric redshift probability distribution functions (PDF) using the {{\small{EAZY}} photometric redshift code \citep{Brammer2008} and stellar masses using an old version of the SED-fitting code {\small{SMPY}} \citep[presented in][]{Duncan2014}. Throughout this work, we use the {\small EAZY} photometric redshifts ($z_{\rm phot}$ and ${\rm PDF}(z)$) computed by \citet{Mundy2017}.} We refer to \citet{Mundy2017} for details on how the photometric redshifts were obtained and detailed comparison to spectroscopic redshift samples. In contrast, we do not use \citet{Mundy2017} stellar-mass estimates but instead our own rederivation of galaxy stellar-masses based on a newer version of {\small{SMPY}} \citep[][]{Duncan2019}. We detail in Sect.~\ref{data:SMpy} how these new stellar-mass estimates are obtained and briefly summarise the main characteristics of the data and of these data products in the three survey regions in Sect.~\ref{data:UDS},~\ref{data:UltraVISTA} and \ref{data:VIDEO}.  Below we give more detail about our derived data products and what we have done to create these and optimise them for our own purposes. 

\subsection{Data Sources}

Our data sources arise from three different fields - the UKIDSS Ultra Deep Survey (UDS), the UltraVISTA survey of the COSMOS field and the deep VIDEO field.  These data sources are part of the REFINE survey for exploring galaxy evolution on deep ground based data.

\subsubsection{UKIDSS Ultra Deep Survey (UDS)}\label{data:UDS}

We use the data aggregated by \citet{Mundy2017} in the UKIDSS Ultra Deep Survey (UDS). It is based on the eighth data release of UDS, the deepest field of the UKIRT Infra-Red Deep Sky Survey \citep[UKIDSS][]{Lawrence2007}, that covers 0.77 deg$^2$ and obtained deep photometry in $J,H,K$ to limiting AB magnitudes of $24.9, 24.2$ and $24.6$ respectively, in 2 arcsec apertures. Complementary observations were aggregated from the CFHT MegaCam $u$, $BVRiz$ from Subaru XMM Deep Survey, $Y$  from ESO VISTA Survey Telescope, IR photometry in four channels from the Spitzer Legacy Program for a combined wavelength range $0.3 < \lambda < 4.6~\mu$m.

The catalogue we use was selected by \citet{Mundy2017} in the $K$ band at $99\%$ completeness $K=24.3$. It contains $\sim 90000$ galaxies out to $z \sim 3.5$ ($90\%$ of galaxies are at $z < 2.4$) in an effective area of $0.63$ deg$^2$.

\citet{Mundy2017} photometric redshifts have a $\sigma_{\rm NMAD} = 0.053 \times (1+z)$ and outlier rate of $\eta = 5\%$ compared to a sample of spectroscopic redshifts. Median $1\sigma$ uncertainties on their rescaled ${\rm PDF}(z)$ is $\sigma_{{\rm PDF}(z)} = 0.040 \times (1+z)$ considering all galaxies $90\%$ complete in stellar-mass.

\subsubsection{COSMOS/UltraVISTA}\label{data:UltraVISTA}

We also use the photometric data aggregated by \citet{Mundy2017} in the COSMOS/UltraVISTA survey. It is based on the publicly available $K_s$ selected catalogue of \citet{Muzzin2013}, that observes the COSMOS field \citep[Cosmological Evolution Survey,][]{Scoville2007} with the ESO Visible and Infrared Survey Telescope for Astronomy (VISTA) telescope and covers a effective area of 1.62 deg$^2$. PSF-matched magnitudes obtained in 2.1 arcsec apertures are provided for 30 bands in the wavelength range $0.15 < \lambda < 24~\mu$m. The catalogue was selected in the VISTA $K_s$ band at $90\%$ completeness magnitude of $K_s = 23.4$ and contains $\sim 150 000$ galaxies out to $z \sim 3$ ($90\%$ of galaxies are at $z < 1.8$).

\citet{Mundy2017} photometric redshifts have a $\sigma_{\rm NMAD} = 0.013 \times (1+z)$ and outlier rate of $\eta = 0.5\%$ compared to a sample of spectroscopic redshifts. Median $1\sigma$ uncertainties on their rescaled ${\rm PDF}(z)$ is $\sigma_{{\rm PDF}(z)} = 0.012 \times (1+z)$ considering all galaxies $90\%$ complete in stellar-mass.

\subsubsection{VIDEO}\label{data:VIDEO}

We use data from the VISTA Deep Extragalactic Observations \citep[VIDEO,][]{Jarvis2012} in the near-IR $Z, Y, J, H, K_s$ bands, matched to the 1 deg$^2$ of the Canada-France-Hawaii Telescope Legacy Survey Deep-1 field (CFHTLS-D1) in optical $u, g, r, i, z$ bands, covering the wavelength range $0.3 < \lambda < 2.1~\mu$m.
We use the VISTA $K_s$ selected catalogue of VIDEO June 2015 release cut at $90\%$ completeness magnitude of $K_s = 22.5$ \citep[see][for details on the completeness simulations]{Mundy2017} and contains $\sim 55 000$ galaxies out to $z \sim 2.5$ ($90\%$ of galaxies are at $z < 1.8$).

\citet{Mundy2017} photometric redshifts of this field have a $\sigma_{\rm NMAD} = 0.044 \times (1+z)$ and outlier rate of $\eta = 2.9\%$ compared to a sample of spectroscopic redshifts. Median $1\sigma$ uncertainties on their rescaled ${\rm PDF}(z)$ is $\sigma_{{\rm PDF}(z)} = 0.034 \times (1+z)$ considering all galaxies $90\%$ complete in stellar-mass.

\subsection{Stellar Masses}\label{data:SMpy}

We obtain individual galaxy stellar mass estimates using the SED-fitting code {\small{SMPY}} in its version presented in \citet{Duncan2019}. We use \citet{BC03} stellar population synthesis models with a \citet{Chabrier2003} Initial Mass Function (IMF). Galaxy model ages are allowed to vary between 10 Myrs and 13.7 Gyrs, sampled at 100 values equally spaced in logarithmic units, enforcing that the galaxy cannot be older than the age of the universe at the redshift under consideration in the fit. Metallicities of 0.005, 0.02, 0.2, 1, 1.75 and 2.5 $Z/Z_\odot$ are considered.  We assume \citet{Calzetti2000} dust attenuation curve with strength in the range $0 \le A_V \le 4$, linearly sampled at 12 values. We adopt exponential $\tau$ models for star formation histories (${\rm SFR} \propto e^{-t/\tau}$) both decreasing (positive $\tau$) and increasing (negative $\tau$) with characteristic time-scales $\vert \tau \vert = 0.25, 0.5, 1, 2.5, 5,$ and $10$ Gyrs, with additional short burst $\tau = 0.05$ and continuous ($\tau \gg 1/H_0$) star-formation models, as in \citet{Duncan2019}. As in \citet{Mundy2017}, we do not include nebular emission. The redshift space is sampled at $dz=0.01$ between $z=0.01$ and $z=5$, in line with photometric redshifts in \citet{Mundy2017}. The simulated fluxes are compared to the total observed flux of our data galaxies using the method described in \citet{Duncan2019}. The observed total flux in each filter is obtained from aperture flux as in \citet{Muzzin2013}.

For each galaxy $i$ and each redshift $z$ on the grid,
{\small{SMPY}} returns  ${\rm PDF}_i(X \vert z)$ with $X$ a desired physical parameter of the galaxy stellar population. Here we are
interested in particular in  ${\rm PDF}_i(M_\star \vert z)$ and ${\rm PDF}_i({\rm sSFR} \vert z)$, where $M_\star$ and ${\rm sSFR}$ are respectively the stellar mass and specific star formation rate of the galaxy. We can then combine these PDFs with the photometric redshift PDF output of {\small{EAZY}} obtained by \citet{Mundy2017} to build : 
\begin{eqnarray}
\begin{aligned}
{\rm PDF}_i(M_\star, z) = {\rm PDF}_i(M_\star \vert z) \times {\rm PDF}_i(z),\\
{\rm PDF}_i({\rm sSFR}, z) = {\rm PDF}_i({\rm sSFR} \vert z) \times {\rm PDF}_i(z).
\end{aligned}
\end{eqnarray}

\begin{figure}
\vspace{0cm} \hspace{-0.25cm}  \includegraphics[width=0.5\textwidth]{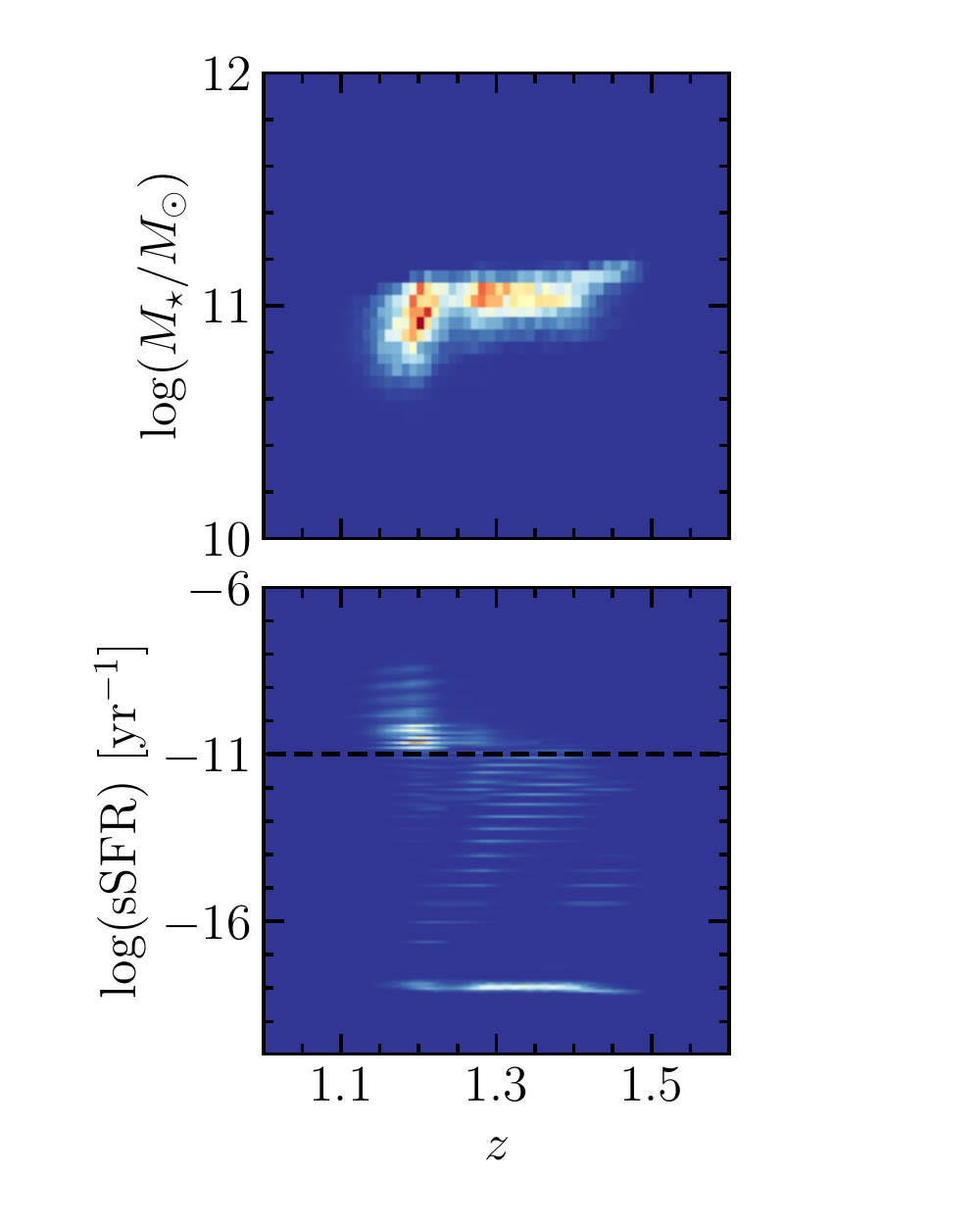}
        \caption{Examples of two-dimensional PDFs for one galaxy in VIDEO. {\it Top}: ${\rm PDF}(M_\star,z)$. {\it Bottom}: ${\rm PDF}({\rm sSFR},z)$. The black dashed line show the sSFR cut used to select quiescent galaxies in Sect.~\ref{sec:fQ} Interestingly, this galaxy exhibits two regions of high probability in the parameter space, that corresponds to similar $z$ and $M_\star$ but very different sSFR.}
\label{fig:2D_PDF_eg}
\end{figure}

\noindent An example of such two-dimensional PDFs are shown in Fig.~\ref{fig:2D_PDF_eg}. 

The general philosophy in our work is to use all the information contained in these delocalised estimates of redshift ($z$), stellar-mass ($M_\star$) and specific star-formation rate (${\rm sSFR}$), whenever possible. In particular, these are used to compute galaxy properties at a given value of any of these parameters. For example, the number of galaxies with stellar-mass $M_{\rm inf} < M_\star < M_{\rm sup}$ in the redshift range  $z_{\rm inf} < z < z_{\rm sup}$ is computed directly from the individual 2D ${\rm PDF}_i(M_\star, z)$ through: 
\begin{equation}
N = \sum_{\substack{i \in\\ {\rm all~galaxies}}} \int_{z_{\rm
    inf}}^{z_{\rm sup}} \int_{M_{\rm inf}}^{M_{\rm sup}} {\rm
  PDF}_i(M_\star, z)~dM_\star~dz.
\end{equation}

In practice, as noted in \citet{LopezSanjuan2017} the probabilistic nature of the PDFs introduces correlations due to galaxies being spread over several bins. This renders binning necessary to account for these correlations and to obtain realistic number counts. \citet{LopezSanjuan2017} found that the optimal bin size is $\Delta z = \frac{z_{\rm sup}-z_{\rm inf}}{2} = 2 \times \langle \sigma_z^{68}\rangle_{\rm median}(z)$. They find the same scaling between bin size and the typical uncertainty for their absolute magnitude parameter, which is comparable in nature to our stellar-mass parameter. This bin size is also coherent with the choice of \citet{CBprobamem}. In this work we will use the $95\%$ confidence interval (which is equal to $2\sigma$ is the normal approximation) as our bin size (see Sect.~\ref{sec:Ncounts}).

We will outline in the relevant sections how this information is used in each particular situation. This treatment of the information output by SED-fitting codes ({\small EAZY} and {\small SMPY}) allows us to statistically study galaxy properties without losing information. We stress that
compared to the more classical approach that consists in using the
best-fitting solution for redshift and stellar mass, and then using a strong cut on these point-estimates to select galaxies in a given range, our method allows us to properly treat uncertainties on galaxy physical parameter estimates (and their covariances).
  
\subsubsection{Useful notations}\label{sec:useful_nota}  
  
Throughout the paper, we will use the notation $\sigma^{\rm CI}_X$, which is the uncertainty on the point
estimate of parameter X that corresponds to the CI confidence interval. For example,  $\sigma^{68}_z$ is the uncertainty on photometric redshifts built from the $68\%$ confidence interval as $\sigma^{68}_z = 0.5 * (z_{u_{68}}-z_{l_{68}})$, where $z_{l_{68}}$ and $z_{u_{68}}$ are respectively the
lower and upper limit of the $68\%$ confidence interval on ${\rm PDF}(z)$.

In several parts of this work (e.g. group detection, group galaxy membership probabilities and galaxy number counts), we need to estimate the typical uncertainty on a given parameter X at a given redshift $z$, stellar-mass $M_\star$ or at a given $(M_\star,z)$. For example we are interested in the typical redshift uncertainty $\sigma^{68}_z$ for galaxies of stellar mass $M_\star$ located at redshift $z$. While taking the median or the mean uncertainty is more common, we work with the $68\%$ percentile of the individual uncertainty distribution, which we found to be better suited to maximise the efficiency of group detection. For clarity, throughout the paper this quantity is quoted using the following notation $\langle \sigma^{68}_z \rangle_{P_{68}}(M_\star,z)$.

We note $\mathcal{N}(\mu, \sigma)$ the normal distribution of mean $\mu$ and standard deviation $\sigma$ . Other distributions are explicitly named.

\subsection{Mock data}\label{data:mock}

To compute the selection function of our cluster finder algorithm and the reliability of our cluster membership assignments we create semi-realistic mock data sets resembling each survey region from cosmological
simulation lightcones.  We test these mocks in the same way we do our own data to see how well we can retrieve known galaxy clusters and groups at a variety of redshifts.

In recent years, some studies used simulated instrumental pipelines
through which cosmological lightcones are observed to obtain mock
observed photometric catalogues. Applying SED-fitting routines to
these mock photometric catalogues then allow them to have realistic
photometric redshifts and photometric physical parameter estimates for a given lightcone and survey \citep[see e.g.][]{Overzier2013,Laigle2019}. Here, out of simplicity and because we need more statistics than is reasonably achievable with these methods, we instead create "semi-realistic" lightcones.

Mock data sets are built
from the 24 public \citet{Henriques15} lightcones based the
Millennium simulation \citet{Springel05} and {\small L-GALAXIES}
semi-analytical model of galaxy formation with BC03 \citep{BC03} single stellar population \citep{Henriques15} which cover
each $1.4$ deg$^2$ up to redshift $z=6$. $h$-dependent quantities, such as distances, halo masses and stellar masses were converted to our chosen value of $h=0.7$ when needed.

We mimic each survey region geometry and add photometric-like noise to the true values of $K$ band magnitude $m_K$, observed redshift $z$  (cosmological redshift + peculiar velocity), and stellar-mass $M_\star$ typical of the photometric uncertainty in the data. In particular, each lightcone galaxy is assigned a two-dimensional ${\rm PDF}(M_\star,z)$
typical of what is found in the data at the galaxy true $(M_\star^{\rm true},z^{\rm true})$ and then has its  true values of $z$ and $M_\star$ shifted according to this PDF. More details on the method used to perform these steps are given in Appendix~\ref{sec:apendix:mock}

In the end, for each of the three surveys, we
have 24 mock survey regions mimicking the data, in terms of geometry,
uncertainty on the magnitude, redshift and stellar-mass estimates and data products (${\rm PDF}(M_\star,z)$ for each galaxy). Adding the area covered by the 24 mocks for each survey region, they cover a total of $14.87, 35.31, 23.49$ deg$^2$ for mock-UDS, mock-UltraVISTA and mock-VIDEO respectively. We note that total number counts in the mocks are higher than in the data by $\sim 15-20\%$, $\sim 1-5\%$ and $\sim 25-30\%$ in the UDS, UltraVISTA and VIDEO survey regions, respectively. This overall excess and discrepancy is larger than the expected cosmic variance ($\sim 5\%$). It can partly be explained by limited angular resolution of the surveys used compared to the lightcone ($\sim 5\%$) that we did not correct for. In any case, the mocks have magnitude, redshift, and stellar-mass distributions that qualitatively agree (similar shapes) with that of the data, and can thus be used to asses the performances of our group finder algorithm.

\section{The \detectifz~algorithm}\label{sec:DETECTIFz}

In this section, we detail how the DElaunay TEssellation ClusTer
IdentiFication with photo-z algorithm (\detectifz) works. The idea
behind this code was to use the Delaunay Tessellation Field Estimator
\citep[DTFE]{Schaap2000} and its scale free nature, often used to
detect cosmic web filaments \citep[e.g.][]{Sousbie2011a}, in a method to
detect galaxy clusters and groups in photometric data. The idea was
also to design an empirical, model free method based
  only on the information contained in galaxy position (sky
  coordinates and photometric redshift) and stellar-mass (and
  the PDF of these parameters), in contrast with existing efficient matched filter algorithm that need to assume a cluster model for detection \citep[e.g. AMICO:][]{Bellagamba2018}. Detecting clusters and groups solely as stellar mass overdensities is particularly interesting at redshift $z > 1.5$ where {\it a priori} knowledge of (proto-)cluster properties is sparse.

Part of the method is inspired by different previous works, in
particular \citet{Cucciati2018,Hung2020} for the Monte Carlo sampling procedure
and \citet{George11}, \citet{CBprobamem} for the probabilistic
membership assignment, as well as the authors' own previous experience
in galaxy cluster detection \citep{Sarron2018}. The python code of \detectifz~will be made public through a dedicated repository\footnote{https://github.com/fsarron/detectifz}.

This section describes the most general version of the
\detectifz~algorithm i.e. when using ${\rm PDF}(M_\star,z)$ as an
input. It should be noted that the algorithm can also be run in different
degraded modes. For example, it accepts as an input data that consists in independent ${\rm PDF}(z)$ and ${\rm PDF}(M_\star)$ (so neglecting covariance) or even point-estimates of $M_\star$ with or without
uncertainty. If no estimate of $M_\star$ is available at all (e.g. too few photometric bands), group detection can also be run using the galaxy number
density only (rather than $M_\star$ density). Note that the less
information included, the more the performance and accuracy presented in this paper may be degraded. In the subsections below we describe how this tool and methodology works in some detail and how it compares to previous detection algorithms for distant clusters.

\subsection{Monte Carlo sampling and density estimation}

To detect galaxy groups, \detectifz~starts by reconstructing the
density field. In this section, we explicitly describe our method for reconstructing the density
field using Monte Carlo (MC) sampling and Delaunay Tessellation Field Estimation (DTFE) in redshift slices. 

\subsubsection{Monte Carlo sampling}\label{DETECTIFz:MC}

To exploit the information encoded in the ${\rm PDF}(M_\star,z)$ in
group detection, we use a Monte Carlo (MC) method. Using the
{\small PINKY} Python package, we draw $N_{\rm mc} = 100$ samples from the
${\rm PDF}(M_\star,z)$ of each galaxy. We then obtain 100 independent
realisations of the data where $z_{\rm phot, mc}$ and $M_{\star,{\rm mc}}$ are fixed to a given value.

For each of the 100 MC catalogues, for group detection, we apply a
cut in stellar mass selecting galaxies more massive than $10^{10} \
{\rm M}_{\sun}$ and above the $90\%$ stellar mass completeness limit at the
galaxy photo-$z$ i.e.
\begin{equation}
    M_\star(z) > \max \left \{ 10^{10} \ {\rm M}_{\sun} , M_\star^{90\%}(z) \right \},
    \label{eq:Mcut}
\end{equation}
\noindent where $M_\star^{90\%}$ is the $90\%$ stellar-mass
completeness limit of the survey at redshift $z$ taken from \citet{Mundy2017}. We note that $M_\star^{90\%} = 10^{10} \ {\rm M}_{\sun}$ at $z = 2.5, 1.6$ and $1.2$ for UDS, UltraVISTA and VIDEO respectively. At higher redshifts, the effective mass cut is the $90\%$ stellar-mass completeness limit $M_\star^{90\%}$.

\subsubsection{Redshift slicing}

The uncertainty on photometric redshifts is usually an order of
magnitude higher than the typical size of galaxy groups and clusters
in redshift space. The signal-to-noise ratio of overdensities is thus enhanced by computing the projected surface density in redshift slices. For this estimated surface density to be an accurate estimate of the underlying surface density, slices need to encompass a representative fraction of the true underlying galaxy population. This is done by taking the 68$^{\rm th}$ percentile of individual redshift uncertainties  $\langle \sigma_z^{68} \rangle_{P_{68}}(z)$ as the half-width of the slice (see Sect.~\ref{sec:useful_nota} for a definition).

Slices are offset from each other by $dz=0.01$ which is the redshift sampling
rate of the PDFs. This offset is smaller than what is usually found in the literature  \citep[e.g.][]{Adam2019} but this enables a better sampling of the redshift space leading to increased precision on cluster redshift and position as well as better completeness for low mass structures. It is important to note that the width of the redshift slices is roughly that of the typical photo-$z$ uncertainty i.e. $0.05 \times (1+z)$ for the UDS survey for example. This is $5-10$ times larger than the offset between the slices, so we are effectively computing a "running" statistic along the redshift dimension.

\subsubsection{2D density field estimation}\label{DETECTIFz:DTFEMC}

Having defined the extent of the redshift slices, for each of the 100 MC realisations, slices are populated with galaxies according to their $z_{{\rm phot, mc}}$.
For each slice $z_i$ and each Monte Carlo realisation $j$ we estimate the 2D
stellar-mass density field $\Sigma_{M_\star}({\rm mc}_j,z_i)$ using a
modified version of
{\small PYDTFE}\footnote{https://github.com/vicbonj/pydtfe}, a Python
implementation of the Delaunay
Tessellation Field Estimator \citep[DTFE][]{Schaap2000}. The
stellar-mass density map is obtained by weighting each galaxy by its
stellar-mass $M_{\star,{\rm mc}_{j}}$ in the DTFE. The DTFE is projected on
a 2D grid with pixel size of $2.88~{\rm arcsec} \times 2.88~{\rm arcsec} \sim 25~{\rm kpc}
\times  25~{\rm kpc~at}~z \sim 2$. 

The mean surface density map in each slice is computed by taking the
logarithmic mean over all MC realisations:
\begin{equation}
   \log \Sigma_{M_\star}^{\rm mc}(z_i) = \frac{1}{N_{\rm mc}} \sum_j
   \log \Sigma_{M_\star}({\rm mc}_j,z_i).
\end{equation}
\noindent The surface density map $\Sigma_{M_\star}^{\rm mc}(z_i)$ is converted to an
overdensity map following \citet{Cucciati2018} :
\begin{equation}
  \log (1 + \delta_{M_\star}) (z_i) = \frac{\Sigma_{M_\star}^{\rm mc}(z_i)}{\langle
  \Sigma_{M_\star}^{\rm mc}(z_i)\rangle},
\end{equation}
\noindent where $\langle \Sigma_{M_\star}^{\rm mc}\rangle =
10^{\mu}e^{2.652\sigma}$ and $\mu, \sigma$ are respectively the mean
and standard deviation of the $3\sigma$ clipped distribution of $\log \Sigma_{M_\star}^{\rm mc}(z_i)$. Finally once we have obtain a density map for each slice $z_i$, we smooth the 3D data cube (ra,dec,z) along the redshift dimension through convolution with a one-dimensional normal distribution of standard deviation $\sigma_z = 0.02$ to remove strong variations between adjacent slices.

\subsection{Overdensity detection}\label{DETECTIFz:detection}

We detect peaks in the 2D density maps in each slice using the {\small PHOTUTILS} Python package \citep{Photutils2019}.
Peaks are detected at a given signal-to-noise ratio ($S/N$) defined as :
\begin{equation}
    S/N = \frac{\log (1 + \delta_{M_\star}^{\rm mc}) - \mu_\delta}{\sigma_\delta},
\end{equation}
\noindent where $\mu_\delta$ and $\sigma_\delta$ are respectively the
mean and standard deviation of the $3\sigma$ clipped distribution of
$\log (1 + \delta_{M_\star}^{\rm mc})$. So in practice peaks are
selected as contiguous sets of pixels of area $A_{\rm det} > \pi
r_{\rm min}^2 $ whose value follows $\log (1 + \delta_{M_\star}^{\rm
  mc}) > \mu_\delta + (S/N)_{\rm min} \sigma_\delta$. In this work we
set $r_{\rm min} = 250 \ {\rm kpc}$ and $(S/N)_{\rm min} = 1.5$.

\subsection{Multiple detection cleaning}\label{DETECTIFz:cleaning}

As the width of the slices is $5-10$ larger than the offset between the slices, the same real overdensity will be detected in several adjacent slices. Thus, it is necessary to to clean for multiple detection of the same overdensity across adjacent slices.

We rank the individual 2D peak detections by $S/N$ and for each identify 2D peak detections at a distance smaller than $r_{\rm clean} = 500$ kpc from it (at the peak detection redshift). We consider all peak detections within $r_{\rm clean}$, and consider that contiguous groups in redshift (no holes between slices) are detections of the same true 3D structure. We thus remove the linked peak detections from the detection catalogue and repeat the above for the next ranked individual 2D peak detection until no detections are left in the catalogue.

After this cleaning process, we are left with a list of aggregate
detections, centred at the peak of the highest $S/N$ detection in the group for both its position on the sky and its redshift. These
aggregate detections form the raw catalogue of galaxy group candidates.

\subsection{Group photometric redshift PDF}\label{DETECTIFz:PDFz}

The group redshift ${\rm PDF}_{\rm group}(z)$ is computed from the mean $\delta_{M_\star}$ at $r <r_{250{\rm kpc}}(z_{\rm group})$ from the group centre. At each $z_i$ where the $S/N > 1.5$, we take ${\rm PDF}(z_i) = \langle \delta_{M_\star}(z_i) \rangle_{r <r_{250{\rm kpc}}(z_{\rm group})}$, otherwise ${\rm PDF}(z_i) = 0$.
The group PDF may have several peaks. We decided in this work not to refine the group catalogue by splitting multiple peak PDFs into different group candidates, as such a refinement process is not straightforward and prone to error. Yet we note that a careful treatment of this step could improve \detectifz~detection of lower-mass galaxy groups with $M_{200} \lesssim 10^{13.5} {\rm M}_{\sun}$.

\subsection{Group size and catalogue refinement}\label{DETECTIFz:R200}

We compute an estimate of the group radius $R_{200c}$. This
radius is defined as the radius of a sphere surrounding a group such that the mean
interior total mass density is $\Delta = 200$ times the critical density of the Universe.

When dealing with galaxy catalogues, we do not have direct access to
the total mass density (dark matter + baryons). We can however compute an estimate of the cluster size $R_{200, M_\star}$ based on an approach similar to that of \citet{Hansen2005}. This method consists in computing the total stellar mass in a given cylindrical volume of radius $r$ around the cluster
centre. Galaxies in this volume will be a mixture of two populations: cluster galaxies and field galaxies. By assuming cluster galaxies are located in a spherical volume of radius
$r$, while field galaxies are located in the cylindrical volume, and
knowing the expected density for field galaxies, we can compute the
density of cluster galaxies in the sphere. We then look for the radius $r$ such that $\Delta = 200$.

For each MC realisation, we compute the total stellar mass summing the stellar masses of galaxies whose projected distance is less than $r$, and redshift is in the $95\%$ confidence interval around the median of the ${\rm PDF}_{\rm group}(z)$. Our estimate of total stellar mass and its uncertainty in the volume are respectively the median and standard deviation over all MC realisation. 
In the same way, we compute the local field $M_\star$ surface density in an annulus $3 < r < 5~{\rm Mpc}$ and the global field $M_\star$ surface density using the full survey. The corresponding areas and volumes are computed by carefully removing masked areas. From all these quantities, we can form the $M_\star$ volumic density contrast:

\begin{equation}
    \delta_r = \frac{\sum\limits_{r} M_\star - \frac{\Omega_r}{\Omega_{\rm loc}} \sum\limits_{\rm loc} M_\star}{\frac{\Omega_r}{\Omega_{\rm FoV}}\sum\limits_{\rm FoV} M_\star} \times \frac{\mathcal{V}_{\rm sphere}}{\mathcal{V}_{\rm cylinder}}
\end{equation}

\noindent where $\Omega_r$ and $\Omega_{\rm loc}$ and $\Omega_{\rm FoV}$ are the surface of a disc of radius $r$, and the annulus of $3 < r < 5~{\rm Mpc}$ used for the local field and full area of the survey respectively. $\mathcal{V}_{\rm sphere}$ is the volume of the sphere of radius $r$ and $\mathcal{V}_{\rm cylinder}$ is the volume of the cylinder. The uncertainty is propagated accordingly. As in \citet{Hansen2005}, the estimate of $R_{200, M_\star}$ is obtained by fitting $\delta_r \pm \Delta\delta_r$ in the region where the $\Delta \sim 200 \Omega_m^{-1}$ density threshold is crossed.

Some candidate groups ($4\%$ in UDS, $12\%$ in UltraVISTA, $7\%$ in VIDEO) never encompass a mean stellar-mass density greater than 200 times the critical density using our estimate. The catalogue is thus refined by discarding those as they are likely spurious detections.

\subsection{Probabilistic membership assignment}\label{DETECTIFz:pmem}

As an additional output, \detectifz~computes the group membership
probability for each galaxy at $d < 2$Mpc from the group centre i.e. the probability that this galaxy is a group galaxy. The probabilistic membership is computed in a Bayesian formalism similar to that developed in \citet{George11} and \citet{CBprobamem}
in which the probability for a galaxy (${\rm gal}$) to be a member of
a given group ($G$) is defined as the posterior probability $P_{\rm mem}
\equiv P({\rm gal} \in G \vert {\rm PDF}_{\rm gal}(M_\star,z),~{\rm PDF}_{\rm group}(z))$. It is obtained by a convolution of the redshift PDFs of galaxies and groups with a prior based on excess number counts in the group region. We note that we use rescaled (i.e. normalised) probabilities as our final estimate, similarly to the method presented in \citet{CBprobamem}. Details of our implementation, including the likelihood definition, our choice of prior and probability rescaling are presented in Appendix~\ref{sec:appendix:pmem}.

\subsection{Group total stellar mass}\label{DETECTIFz:totmass}

We then use the membership probabilities to compute two total mass proxies for our detected groups. The richness is taken as the number of galaxies inside the group $R_{200,M_\star}$ with a stellar-mass estimate at the group redshift higher than $10^{10.5} {\rm M}_{\sun}$. It is computed from the membership probabilities as follows:
\begin{equation}
    N_{\rm gal} = \sum_{r < R_{200,M_\star}} P_{\rm mem}.
\end{equation}
In the same way, we can compute the group total stellar mass :
\begin{equation}
   \mu_\star = \sum_{r < R_{200,M_\star}} P_{\rm mem} \times M_\star^{\rm gal}(z_{\rm group}).
\end{equation}

With this definition of $\mu_\star$, it may happen that some groups have $\log \mu_\star < 0$. We do a final cleaning of our catalogue discarding those to form our final \detectifz~candidate group catalogue.  When applying ~\detectifz~to our data we find that this is the case for 13 groups in UltraVISTA ($0.9 \%$) and 4 groups in VIDEO ($0.7\%$). We also checked that none of our detected groups have nonphysically high total stellar masses.

\begin{figure*}
    \includegraphics[width=1\textwidth]{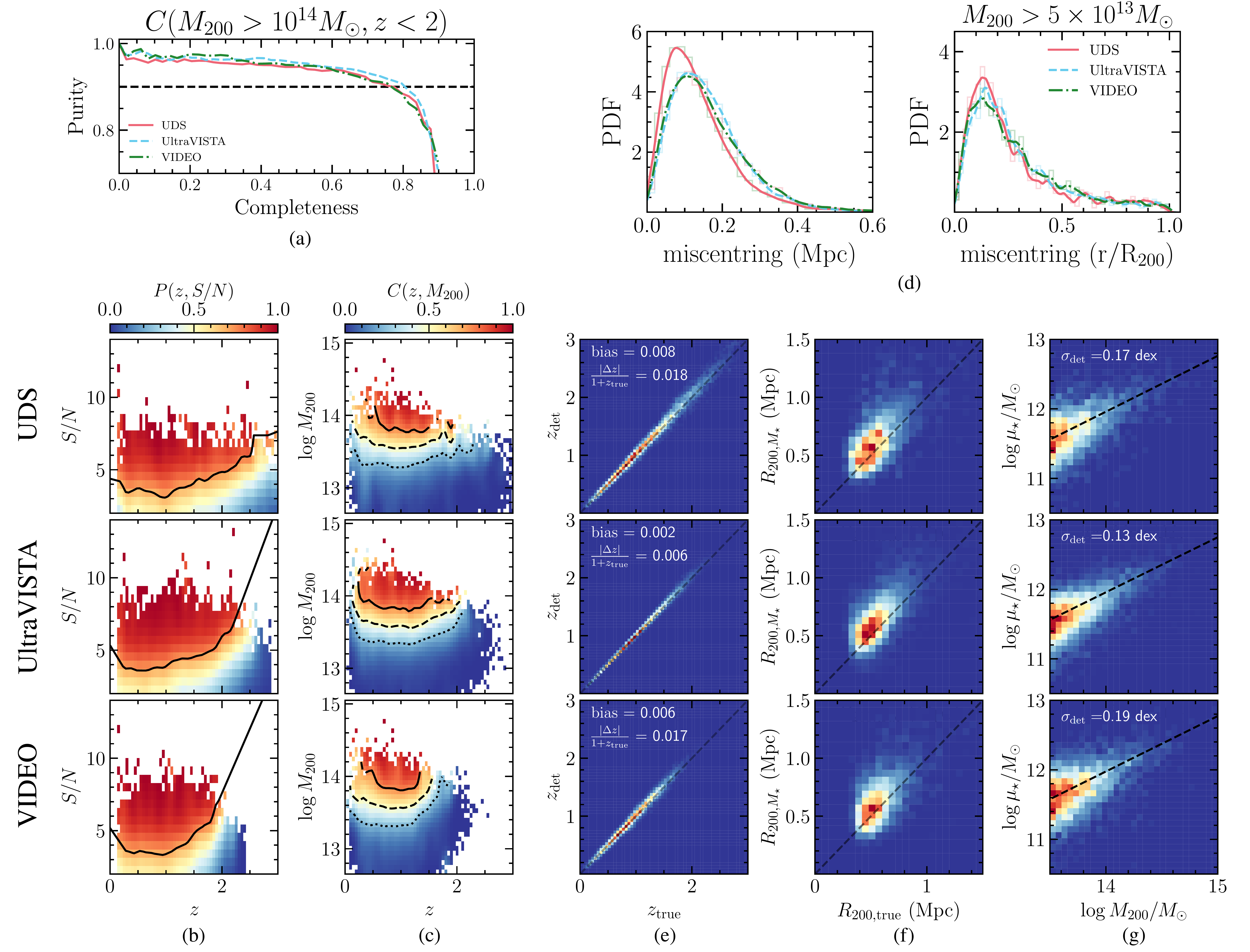}
        \caption{Overview of \detectifz~selection function using several metrics. (a) Purity vs Completeness for galaxy clusters at $z < 2$ in the three survey regions (solid red line for UDS, blue dashed line for UltraVISTA and green dash-dotted line for VIDEO). (b) Purity $P(z,S/N)$ as a function of redshift and signal-to-noise ratio. The raw two-dimensional histogram is computed in bins of size $\Delta z = 0.05$ and $\Delta S/N = 0.75$, but was smoothed with a Gaussian kernel of width $\sigma_z=0.25$ and $\sigma_{S/N} = 0.8$ for display purposes. The black solid line shows the $(S/N)_{75\%}(z) \equiv P(z,S/N) = 75\%$ threshold (see text for details). (c) Completeness $C(z,M_H)$ as a function of redshift and halo mass $M_{200}$, with the $(S/N)_{75\%}(z)$ cut applied that ensures a sample purity $>90\%$. The raw two-dimensional histogram is computed in bins of size $\Delta z = 0.05$ and $\Delta \log M_{200} = 0.08$ but was smoothed with a Gaussian kernel of width $\sigma_z=0.25$ and $\sigma_{\log M_{200}} = 0.20$ for display purposes. Solid, dashed and dotted line show the 75, 50 and 25$\%$ completeness limit respectively. (d) Distribution of group miscentring in Mpc for all matched groups (left) and in units of $r/R_{200}$ for matched groups more massive than $10^{13.5} {\rm M}_{\sun}$ (right). (e) Redshift estimated by \detectifz~$z_{\rm det}$ vs halo redshift $z_{\rm true}$ for matched groups. The bias and scatter is indicated on the top left for each survey region. The dashed line shows the one-to-one line. (f) Estimated group radius $R_{200, M_\star}$ vs true group radius $R_{200, {\rm true}}$. The dashed line shows the one-to-one line. (g) Estimated total stellar mass $\log \mu_\star$ vs halo mass $\log M_{200}$. The dashed line shows the best linear fit to the log-log relation. The scatter around the best-fitting} relation $\sigma_{\rm det}$ is shown on the top left.
\label{fig:C2}
\label{fig:SeF}
\end{figure*} 
\begin{figure*}
    \includegraphics[width=1.0\textwidth]{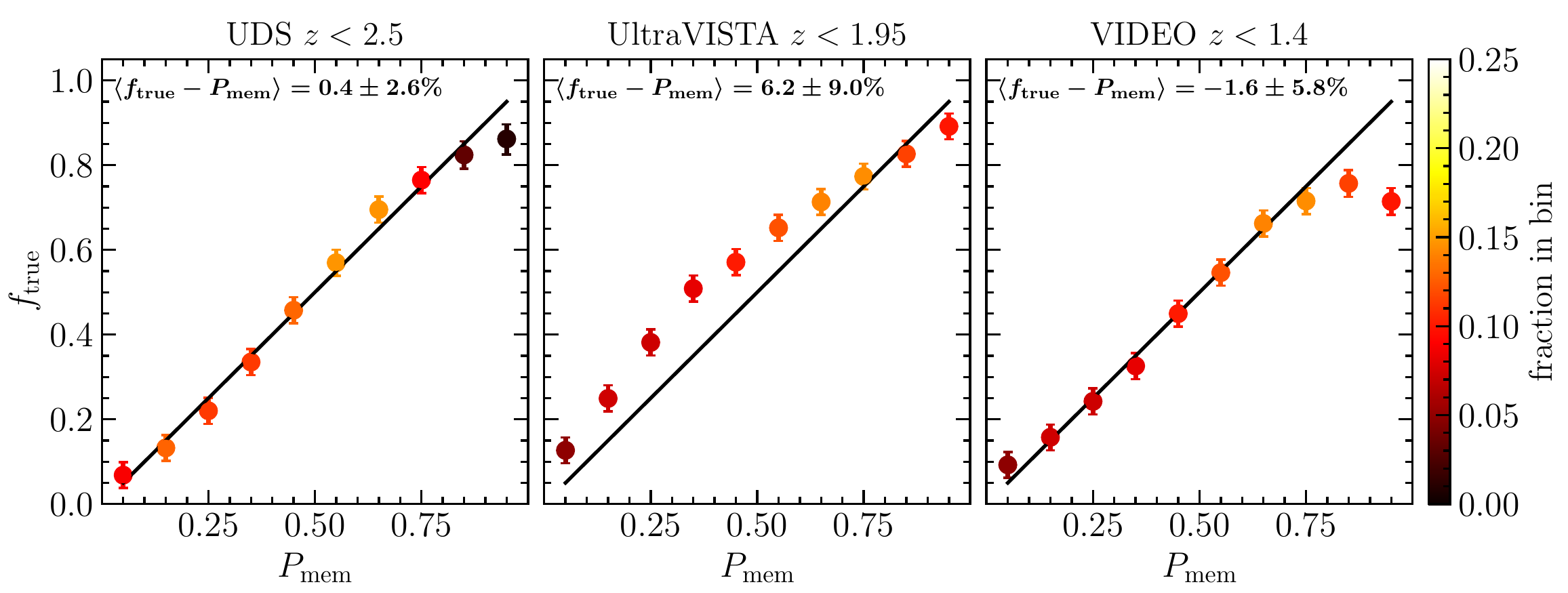}
        \caption{Fraction of true members $f_{\rm true}$ vs the estimated probability of membership $P_{\rm mem}$ for galaxies more massive than $10^{10.25} {\rm M}_{\sun}$ in groups that were match to a halo with mass $M_{200} > 5 \times 10^{13} {\rm M}_{\sun}$. For each matched group, only galaxies located in the matched halo projected $R_{200}$ are considered. Only groups at redshifts $z < 2.5$ where the 90\% stellar-mass completeness limit is greater than $10^{10.25} {\rm M}_{\sun}$ are considered i.e. $z < 2.5$ for UDS, $z < 1.95$ for UltraVISTA and $z < 1.4$ for VIDEO. The bias, taken as the median $f_{true} - P_{\rm mem}$ weighted by the number of galaxies in each bin of $P_{\rm mem}$ is shown on the top right.}
\label{fig:C3}
\label{fig:SeF_Pmem}
\end{figure*}

\section{\detectifz~catalogue}\label{DETECTIFz_cat}
\subsection{Selection function}\label{sec:SelFun}
To compute the selection function of the \detectifz~algorithm, we use the mock data presented in Sect.~\ref{data:mock}. These mock data were constructed from cosmological lightcones and adjusted to be representative of the different surveys considered in this study. As the true properties of galaxies and dark matter haloes they belong to are known in the mocks, we can use these to estimate the performances of the \detectifz~algorithm for each of the three survey regions we examine. 

To match true haloes and detected groups, we use the rank matching method described in \citet{Adam2019}. We rank detected cluster by decreasing $S/N$ and dark matter haloes by decreasing halo mass $M_{200}$. We go down the list of ranked haloes and for each look for matches at sky separation smaller than the halo $R_{200}$ and with a redshift within $\pm 2 \times \langle \sigma_z^{68}\rangle_{P_{68}}(z)$ from the halo true redshift. The halo centre in (RA,Dec) and redshift are computed using the stellar-mass weighted average of true halo members more massive than $M_\star > 10^{10} {\rm M}_{\sun}$. We match groups detected at $S/N > 1.5$ with haloes of mass $M_{200} > 10^{12.5}{\rm M}_{\sun}$ that have at least 3 member galaxies with $M_\star > 10^{10}{\rm M}_{\sun}$, as these haloes are detectable by our algorithm {\it a priori}.\\
\indent The selection function quantifies the ability of our group finder algorithm to detect galaxy groups and clusters and how contaminated it is by false detections. This can be expressed through two quantities, the completeness and the purity. The completeness $C$ is the fraction of true haloes above a given mass $M_{200}$ that are detected by our algorithm. The purity $P$ is the fraction of our detections that are actual galaxy groups ($P = 1 -$ false detection rate).\\  
\indent In Fig.~\ref{fig:SeF}\textcolor{blue}{a}, we plot, using our method on these mocks, the purity of our sample versus the completeness for haloes of $M_{200} > 10^{14}{\rm M}_{\sun}$ (i.e. galaxy clusters) and $z < 2$. At the $90\%$ purity level, we reach a completeness of $\sim 80\%$ for clusters in all three fields. When considering the full catalogue at $S/N > 1.5$, we reach a completeness $>90\%$ for halo mass $M_{200} > 10^{14}{\rm M}_{\sun}$, and $z < 2.5$ for a purity of $\sim 55\%$ in the three fields.\\
\indent To have a more detailed view of our selection function, the purity can be expressed as a function of redshift $z$ and detection signal-to-noise ratio $S/N$ :
\begin{equation}
P(z,S/N) = \frac{N_{\rm match}(z,S/N)}{N_{\rm det}(z,S/N)}.
\end{equation}
\noindent where $N_{\rm match}$ is the number of detections matched to haloes in the simulation and $N_{\rm det}$ the number of detections. The completeness can be expressed as a function of redshift $z$ and halo mass $M_{200}$ : 
\begin{equation}
C(z,M_{200}) = \frac{N_{\rm match}(z,M_{200})}{N_{\rm true}(z,M_{200})},
\end{equation}
\noindent where $N_{\rm true}$ the number of haloes in the simulation.

$P(z,S/N)$ is shown in Fig.~\ref{fig:SeF}\textcolor{blue}{b}. The purity is an increasing function of $S/N$ at all redshift (higher $S/N$ imply higher purity), and shows a dependence with redshift that makes higher $S/N$ detections having a higher probability to be false detections at higher redshift. This can partially be explained by the fact our stellar-mass cut starts to increase at redshifts where the stellar-mass completeness kicks in (see Sect.~\ref{DETECTIFz:MC}). This behaviour motivated us to define a redshift dependent $S/N$ cut in order to select a sample with high purity, that we will use in the rest of the paper. This cut, $(S/N)_{75\%}(z)$, is chosen as the $S/N$ at which the purity is $75\%$ at each redshift. Using this cut we select a sample that is $\ge 90\%$ pure at all redshifts.

In Fig.~\ref{fig:SeF}\textcolor{blue}{c}, we show the completeness $C(z,M_{200})$ with the $(S/N)_{75\%}(z)$ cut in purity applied. This means the completeness we show is for a sample that is $>90\%$ pure at all redshift. For galaxy clusters ($M_{200} > 10^{14}{\rm M}_{\sun}$), the completeness is $83\%, 84\%$ and $78\%$ at $z < 2$ for UDS, UltraVISTA and VIDEO respectively. For intermediate mass groups ($5 \times 10^{13} {\rm M}_{\sun} < M_{200} \le 10^{14} {\rm M}_{\sun}$), it is $\sim 65\%$ at redshift $z < 1$ in all three fields and $65\%, 60\%$ and $53\%$ for UDS, UltraVISTA and VIDEO respectively at redshift $1 < z < 2$. This differences between the three fields is a direct consequence of the different depth (in terms of observed flux) of the three fields. In this range of halo mass, the completeness at $2 < z < 3$ is $23\%$ in UDS and $16\%$ in UltraVISTA, while there are no detections at these redshifts in VIDEO.

Using our detected groups that are matched to true haloes, we compute statistics about our estimate of group properties. Fig.~\ref{fig:SeF}\textcolor{blue}{d} shows the distribution of group miscentring, defined as the distance between \detectifz~centre and halo centre on the sky, in Mpc (left) and in units of $r/R_{200}$ (right). We find that $75\%$ of \detectifz~groups are miscentred by less than $185, 215$ and $220$ kpc for UDS, UltraVISTA and VIDEO respectively. If looking at groups with $M_{200} > 5 \times 10^{13} {\rm M}_{\sun}$, $75\%$ of \detectifz~are miscentred by less than $0.35 \times R_{200}$ for all three fields. 

In Fig.~\ref{fig:SeF}\textcolor{blue}{e}, we show the distribution of detected redshifts versus true redshifts. The redshift recovered by \detectifz~has a very small bias $<0.009$ overall, and $<0.005$ at redshift $z < 1.5$. The median scatter $\vert z_{\rm true} - z_{\rm det}\vert / (1+z_{\rm true})$ is $0.018, 0.006$ and $0.017$ in UDS, UltraVISTA and VIDEO respectively i.e. half the typical uncertainty on the redshift of individual galaxies.

In Fig.~\ref{fig:SeF}\textcolor{blue}{f}, we show the distribution of our estimate of the group radius $R_{200,M_\star}$ versus the true $R_{200}$ of haloes (estimated from the dark matter density contrast). Our estimate is slightly over-estimated by $11\%, 10\%$ and $7\%$ for groups with mass $5 \times 10^{13} {\rm M}_{\sun} < M_{200} < 2 \times 10^{14}{\rm M}_{\sun}$ within the UDS, UltraVISTA and VIDEO fields respectively. At $M_{200} > 2 \times 10^{14} {\rm M}_{\sun}$ this bias falls to $1\%, 2\%$ and $0.5\%$ within the UDS, UltraVISTA and VIDEO fields, respectively. The typical  scatter on the estimate is $\sim 0.15$ Mpc at all group mass.

Finally, Fig.~\ref{fig:SeF}\textcolor{blue}{g} shows the distribution of our mass proxy $\log \mu_\star$ formed by summing the stellar masses of galaxies in $R_{200,M_\star}$, weighted by their probability of memberships $P_{\rm mem}$ (see Sect.~\ref{DETECTIFz:totmass}). This estimator is well correlated with halo mass $M_{200}$. We fit the relation between both variables using a linear model in log-log space using only haloes with mass $M_{200} > 10^{13.5}$. The median measured scatter around the best-fitting relation is $\sigma_{\rm meas} = 0.16, 0.15$ and $0.16$ dex for UDS, UltraVISTA and VIDEO respectively. As argued in \citet{Adam2019}, to compare different mass proxies that scale differently with halo mass and to account for the intrinsic scatter in the total stellar mass versus halo mass relation, one can form the quantity $\sigma_{\rm det} = \sqrt{\sigma_{\rm meas}^2/s_{\rm meas}^2 - \sigma_{\rm int}^2/s_{\rm int}^2}$, which is the scatter due to the detection process.  This equation is such that $\sigma_{\rm int}$ is the intrinsic scatter, and $s_{\rm meas}^2$ and $s_{\rm int}^2$ are the slopes of the scaling relations for the measured total stellar mass and true total stellar mass, respectively. Using this we find: $\sigma_{\rm det} = 0.17, 0.13$ and $0.19$ respectively. In particular, for groups with mass $M_{200} > 10^{14}{\rm M}_{\sun}$ and $z < 2$, we find $\sigma_{\rm det} = 0.16$ dex for the UDS field, a value directly comparable to the ones given in Table 2 of \citet{Adam2019} and which are competitive with the best richness estimates they presented.

The results presented in the previous paragraph show that globally, summing our membership probabilities gives a good mass proxy. To have a more detailed view, we can verify how accurate our probability of membership $P_{\rm mem}$ is for individual galaxies by comparing it to the fraction of true group members. Only members of groups matched to a true halo and located at a distance less than $R_{200,{\rm true}}$ of the cluster centre are used in the analysis. In Fig~\ref{fig:SeF_Pmem} we show the fraction of galaxies that are true cluster members in bins of $P_{\rm mem}$ for each survey region. We find an overall good agreement, with a deviation of less than $10\%$ in all bins, showing the strength of our probabilistic membership assignment. 

\subsection{The REFINE group catalogue}

We applied the \detectifz~algorithm to the UDS, COSMOS/UltraVISTA and CFHTLS-D1/VIDEO survey regions. We detect respectively $540, 1495$ and $553$ candidate groups and clusters at $S/N > 1.5$. We have shown in Sect.~\ref{sec:SelFun} that it is necessary to apply a stricter cut in $S/N$ to lower the false detection rate (higher purity). For the remaining of the paper, we use a very strict $(S/N)_{75\%}(z)$ that guarantees $75\%$ purity for clusters with $S/N = (S/N)_{75\%}$ at all redshift as illustrated in Fig.~\ref{fig:SeF}(b). With this cut, the mean purity is $>90\%$ at all redshifts. While this impacts the completeness of the sample, high purity is preferred for our study of the group galaxy quenched fraction in Sect.~\ref{sec:fQ}. The estimated completeness as a function of redshift $z$ and halo mass $M_{200}$ is shown in Fig.~\ref{fig:SeF}\textcolor{blue}{c}. At this high purity we have a sample of 448 galaxy groups up to $z = 2.5$ (77 in UDS, 255 in UltraVISTA and 116 in VIDEO) with 53 groups at $z > 1.5$ (14 in UDS, 30 in UltraVISTA and 9 in VIDEO).

The redshift, $S/N$ and $\mu_\star$ distributions of our group samples in each of the three survey regions and for the three survey regions combined (REFINE) are shown in Fig~\ref{fig:clusproperties}. Dashed lines show the distributions considering the full sample ($S/N > 1.5$, completeness $>90\%$ and purity $\sim 55\%$), the solid lines show the distributions for the pure sample ($S/N > (S/N)_{75\%}(z)$, completeness $\sim80\%$ and purity $> 90\%$). Three colour $iJK$ images and some properties of probable group members for three groups at different redshifts in the UDS survey region are presented in Appendix~\ref{sec:appendix:iJK}.

\subsection{Comparison to existing cluster catalogues}\label{sec:match}

To further assess the quality of our group sample and quantify which of our detections are new, we compared our group catalogues to the literature. In the three survey regions considered in this work, many previous studies have tackled group or cluster detection, in a variety of redshift ranges, using either optical and near-infrared data or X-ray data. 

\begin{figure*}
\centering
    \includegraphics[width=1\textwidth]{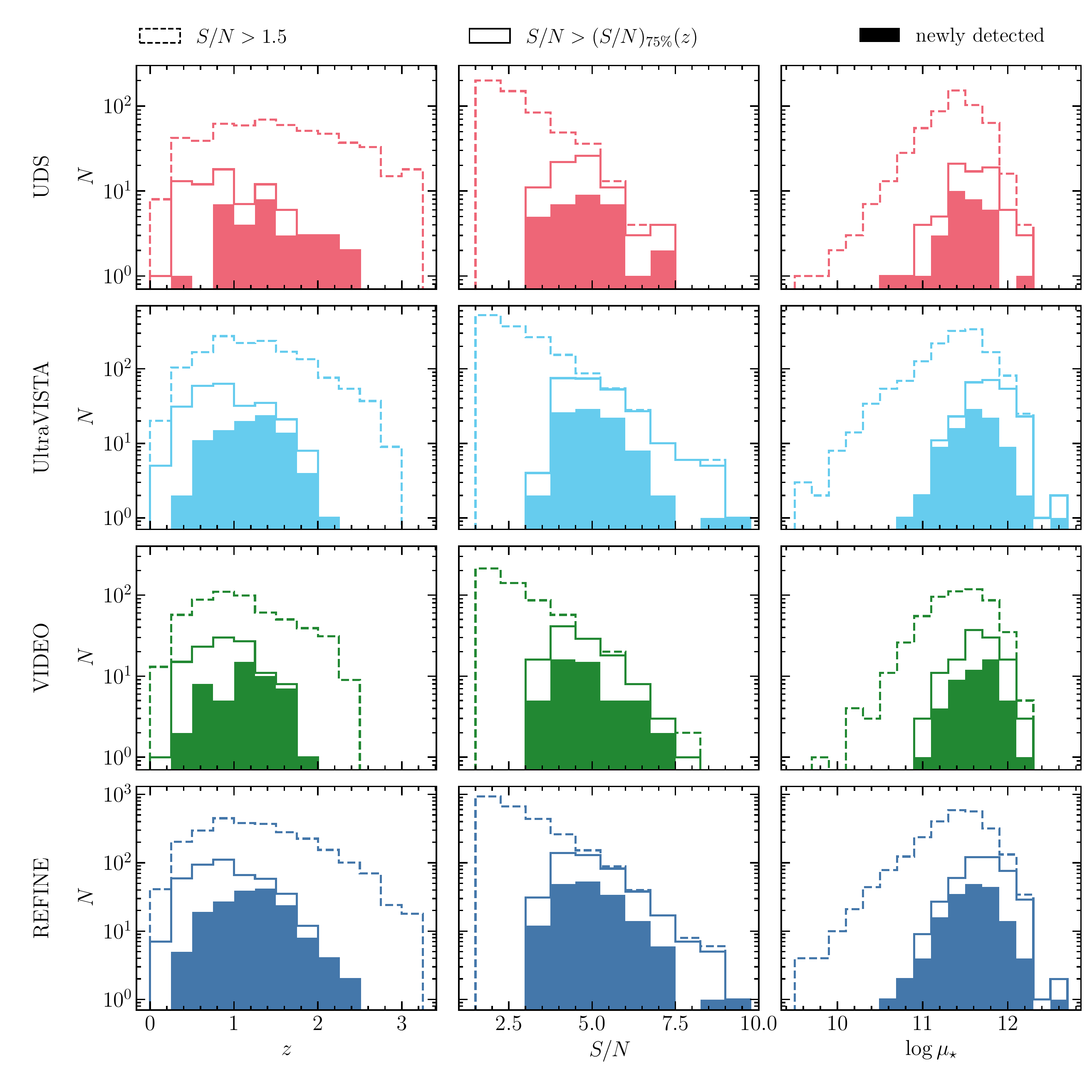}
        \caption{Left: Histogram of redshifts of REFINE candidate groups. Middle: Histogram of signal-to-noise ratio $S/N$ of REFINE candidate groups. Right: Histogram of total stellar mass $\log \mu_\star$ of REFINE groups. From top to bottom the rows display the UDS, UltraVISTA, VIDEO and REFINE (UDS+UltraVISTA+VIDEO) survey. We show the distributions  of all candidate groups ($S/N > 1.5$, dashed line), $90\%$ pure group sample ($S/N > (S/N)_{75\%}(z)$, solid line) and the newly discovered groups (filled histogram).}
\label{fig:clusproperties}
\end{figure*}

To match our clusters to other catalogues, we used two-way geometrical matching \citep[see][]{Adam2019}, considering two detections are matched if they are at a distance $d < 1$ Mpc on the sky (at the redshift of the \detectifz~group) and $\vert \Delta z \vert < 2\times \langle\sigma_z^{68}\rangle_{P_{68}}(z)$. As catalogues from the literature have different completeness and purity level, we cannot properly assess our own completeness and purity from comparison to these. Yet for any given catalogue from the literature, we compute the percentage of their clusters located in the same field as ours that we detect at $S/N > 1.5$ (that we call re-detected groups) and the number of \detectifz~secure detections ($S/N > (S/N)_{75\%}(z)$) that are not detected in other surveys (that we call newly detected groups). Overall, in our pure sample, we find  170 newly detected galaxy groups (31 in UDS, 91 in UltraVISTA and 48 in VIDEO). At $z > 1.5$, 38 groups of our pure sample are newly detected (11 in UDS, 19 in UltraVISTA and 8 in VIDEO). Properties of the newly detected clusters for the full REFINE survey are shown in the bottom row of Fig.~\ref{fig:clusproperties} as filled histograms.  We discuss these newly detected clusters in the subsections below.

\subsubsection{UDS}
In the UDS field, we matched our group catalogue to public catalogues from the literature. \citet{Lee2015}, \citet{Socolovsky2018} and \citet{Galametz2018} detected 46, 39 and 34 galaxy group and clusters in the UDS field of view using data from the UDS survey in the redshift range $0.5 < z < 2$, $0.5 < z < 1$ and $0.6 < z < 0.7$ respectively. The UDS field is also located inside several other surveys in which clusters and groups were detected. We list them here with the redshift ranges of the clusters. In the W1 field of the CFHTLS survey we used catalogues from \citet{Ford+15} (118 clusters at $0.2 < z < 1$ in UDS), \citet{RedGold} (3 clusters at $0.45 < z < 0.79$ in UDS) and \citet{Sarron2018} (5 clusters at $0.53 < z < 0.7$ in UDS). In the HSC-SSPxunWISE  survey we used the catalogue of \citet{Wen2021} (12 clusters at $0.40 < z < 1.43$ in UDS). In the X-rays, we used the catalogues of \citet{Finoguenov2010} (43 clusters at $0.19 < z < 2.15$ in UDS) from the Subaru-XMM Deep Field (SXDF) and \citet{Adami2018} (10 clusters at $0.18 < z < 1.1$ in UDS) from the XXL survey.

Results from matching these catalogues to ours are summarised in Table~\ref{tab:matchUDS}, and are shown in the first row of Fig.~\ref{fig:clusproperties} as filled histograms. We also present a comparison between our group mass proxy $\mu_\star$ and those of other catalogues in online Appendix~\textcolor{blue}{D}.

\begin{table}
	\centering
	\caption{The number ($N_{\rm match}$) of groups in catalogues from the literature also detected by \detectifz, and the percentage of the original catalogue re-detected in the UDS field.}
	\label{tab:matchUDS}
	\begin{tabular}{lccr} 
		\hline
		Reference & $N_{\rm match}$ & \% re-detected \\
		\hline
		\citet{Lee2015} & 39 & 85\%\\
		\citet{Socolovsky2018} & 31 & 80\%\\
		\citet{Ford+15} & 71 & 60\%\\
		\citet{RedGold} & 1 & 33\%\\
		\citet{Sarron2018} & 4 & 80\%\\
		\citet{Wen2021} & 10 & 83\%\\
		\citet{Finoguenov2010} & 32 & 74\%\\
		\citet{Adami2018} & 7 & 70\%\\
		\citet{Galametz2018} & 17 & 50\%\\
		\hline
	\end{tabular}
\end{table}

\subsubsection{COSMOS/UltraVISTA}

In the UltraVISTA field, we matched our group catalogue to six public catalogues from the literature obtained from optical and X-ray data. In the optical, \citet{Bellagamba2011} detected 142 clusters in the redshift range $0.12 < z < 0.8$ in the UltraVISTA survey region, while \citet{Wen2011} detected 209 clusters in the redshift range $0.17 < z < 1.61$. \citet{Chiang2014} looked for overdensities of galaxies based on photometric redshifts at $z > 1.5$ on scales of $\sim 15$ Mpc. They detected 36 such overdensities in the range $1.6 < z < 3.09$. Recently, \citet{Ando2020} looked for protocluster cores at $z > 1.5$ using galaxy photometric redshifts. They detected 75 protocluster cores in the range $1.5 < z < 2.93$. \citet{Wen2021} looked for galaxy clusters in the HSC-SSPxunWISE field using optical and near-infrared data. In the area of the COSMOS/UltraVISTA, they detected 69 galaxy clusters in the redshift range $0.2 < z < 1.51$. In the X-rays, \cite{Gozaliasl2019} detected 230 galaxy clusters in the COSMOS/UltraVISTA field at redshift $0.05 < z < 1.53$.

Results from matching these catalogues to ours are summarised in Table~\ref{tab:matchUltraVISTA}, and are shown in the second row of Fig.~\ref{fig:clusproperties} as filled histograms. We also present a comparison between our group mass proxy $\mu_\star$ and those of other catalogues in online Appendix~\textcolor{blue}{D}.

\begin{table}
	\centering
	\caption{The number ($N_{\rm match}$) of groups in catalogues from the literature also detected by \detectifz, and the percentage of the original catalogue re-detected in the COSMOS/UltraVISTA field.}
	\label{tab:matchUltraVISTA}
	\begin{tabular}{lccr} 
		\hline
		Reference & $N_{\rm match}$ & \% re-detected \\
		\hline
		\citet{Chiang2014} & 8 & 22\%\\
		\citet{Ando2020} & 46 & 61\%\\
		\citet{Gozaliasl2019} & 144 & 63\%\\
		\citet{Bellagamba2011} & 115 & 81\%\\
		\citet{Wen2011} & 157 & 75\%\\
		\citet{Wen2021} & 49 & 71\%\\
		\hline
	\end{tabular}
\end{table}

\subsubsection{CFHTLS-D1/VIDEO}

In the VIDEO field, we matched our group catalogue to seven public catalogues from the literature obtained from optical and X-ray data in different survey areas.
As this field is located in the W1 field of the CFHTLS, as for UDS we used the catalogues of \citet{Ford+15} (152 clusters at $0.2 < z < 1.01$ in CFHTLS-D1), \citet{RedGold} (15 clusters at $0.14 < z < 1.02$ in CFHTLS-D1) and \citet{Sarron2018} (6 clusters at $0.18 < z < 0.53$ in CFHTLS-D1).  In the HSC-SSPxunWISE survey we used the catalogue of \citet{Wen2021} (27 clusters at $0.25 < z < 1.53$ in CFHTLS-D1). In the X-rays, we used the catalogue of \citet{Gozaliasl} (42 clusters at $0.14 < z < 1.08$ in CFHTLS-D1) and two catalogues from the XXL survey, \citet{Adami2018} (10 clusters at $0.04 < z < 1.06$ in CFHTLS-D1) and \citet{Trudeau2020} that detected high redshift XXL clusters counterparts in the optical/near-infrared (8 clusters at $0.82 < z < 1.94$ in CFHTLS-D1).

Results from matching these catalogues to ours are summarised in Table~\ref{tab:matchVIDEO} and shown in the third row of Fig.~\ref{fig:clusproperties} as filled histograms. We also present a comparison between our group mass proxy $\mu_\star$ and those of other catalogues in online Appendix~\textcolor{blue}{D}.

\begin{table}
	\centering
	\caption{The number ($N_{\rm match}$) of groups in catalogues from the literature also detected by \detectifz, and percentage of the original catalogue re-detected in the CFHTLS-D1/VIDEO field.}
	\label{tab:matchVIDEO}
	\begin{tabular}{lccr} 
		\hline
		Reference & $N_{\rm match}$ & \% re-detected \\
		\hline
        \citet{Gozaliasl} & 30 & 65\%\\
		\citet{Ford+15} & 117 & 77\%\\
		\citet{RedGold} & 14 & 93\%\\
		\citet{Adami2018} & 8 & 80\%\\
		\citet{Sarron2018} & 6 & 100\%\\
		\citet{Wen2021} & 26 & 96\%\\
        \citet{Trudeau2020} & 8 & 89\%\\
		\hline
	\end{tabular}
\end{table}

\section{Quenched fractions}\label{sec:fQ}

This section is dedicated to a preliminary study of the cluster/group galaxy quenched fraction in our candidate cluster/group sample. We voluntarily limit ourselves here to the study of galaxies with stellar masses $10^{10.25} < M_\star/{\rm M}_{\sun} < 10^{11}$ and within $0.5 \times R_{200}$ the group centre, in groups with total stellar mass $\log \mu_\star > 11.25$. The lower stellar mass limit ensures $90\%$ stellar-mass completeness for all groups. Using these cuts, we are left with 403 galaxy groups in the range $0.12 \le z < 2.32$ for which the estimated purity is $> 90\%$. The upper mass limit removes the contribution of massive galaxies for which mass quenching is thought to be the dominant quenching process. It also allows us to remove the relative excess of these galaxies in groups compared to the field that would prevent any interpretation of the results in terms of the effect of the environment. Overall, we checked that using this mass range ensures that we are comparing samples with similar stellar-mass distributions and mean galaxy stellar mass $M_\star \sim 10^{10.6} {\rm M}_{\sun}$ in the groups and $M_\star \sim 10^{10.55} {\rm M}_{\sun}$ in the field. The radial limit means that we are only probing the group core that is characterised by a higher density contrast. Doing this makes it easier to isolate the effect of the group on the quenching fraction. Further investigation of the stellar-mass and radial dependence through computation of group galaxy stellar mass function (GSMF) and radial distributions respectively will be presented in a future work.

\subsection{Number counts}\label{sec:Ncounts}

To compute the group galaxy quenched fraction, we first compute the galaxy number counts for {\it all} galaxies and {\it quenched} galaxies in groups and outside groups (field) respectively. When considering galaxies at all sSFR values, number counts are computed using the method introduced Sect.~\ref{data:SMpy} with appropriate binning in redshift and stellar mass.

The final quantity we want to estimate is the number of galaxies with a stellar mass in the range $10^{10.25} < M_\star / {\rm M}_{\sun} < 10^{11}$ at the redshift of the group, taking into account the uncertainty we have on this quantity that is encoded in the ${\rm PDF}_{\rm group}(z)$. This requires smoothing the raw number counts in the redshift dimension and then taking the mean number counts with ${\rm PDF}_{\rm group}(z)$ as a probability measure. In practice, starting from the ${\rm PDF}(M_\star,z)$ of each galaxy, we first compute the raw (unbinned) number counts as : 
\begin{equation}
N(M_\star,z) = \sum_{\substack{i_{\rm gal}}} {\rm
  PDF}_{i_{\rm gal}}(M_\star, z).
\end{equation}
\noindent Number counts smoothed (binned) in the redshift direction are obtained at each $(M_\star,z)$ by summing over adjacent bins with $z \in [z-\Delta z,z+\Delta z]$: 
\begin{equation}
N^{\rm binned}(M_\star,z) = \int_{z-\Delta z}^{z+\Delta z} N(M_\star,z) dz, 
\label{eq:Nbinned}
\end{equation}
where $\Delta z = \langle \sigma_z^{95} \rangle_{P_{68}} (M_\star,z)$. From these binned number counts we can then compute our estimate of number counts at the group redshift through : 
\begin{equation}
N_{z_{\rm group}} = \int_{10^{10.25}}^{10^{11}} \int_{0}^{5} N^{\rm binned}(M_\star,z)~{\rm PDF}_{\rm group}(z)~dz~dM_\star.
\label{eq:Ngroup}
\end{equation}\\

\noindent The quenched number counts at the group redshift are computed in the same way, but adding our knowledge about each galaxy sSFR encoded in ${\rm PDF}({\rm sSFR},z)$. From this PDF, we can compute for each galaxy its probability to be quenched at redshift $z$. Galaxies are considered to be quenched if they have ${\rm {\rm sSFR}} < 10^{-11} {\rm yr}^{-1}$. We use the same sSFR limit at all redshifts in line with some previous works \citep[e.g.][]{Ilbert2010}. We discuss the implication of this cut compared to some other choices in the literature in Sect.~\ref{sec:Discussion_sSFR}. From the ${\rm PDF}({\rm sSFR},z)$, we compute the probability that the galaxy is quenched at each redshift $z$, by integrating the ${\rm PDF}({\rm sSFR},z)$ up to our chosen sSFR limit: 
\begin{equation}
P^q(z) = \frac{\int_{-\infty}^{10^{-11}} {\rm PDF}({\rm sSFR},z)~d{\rm sSFR}}{\int{\rm PDF}({\rm sSFR},z)~d{\rm sSFR}},
\end{equation}
\noindent such that at a given redshift $z$, $0 < P^q(z) < 1$ and the total probability (quenched and non-quenched) at a each redshift sums to one. This allows to get an estimate of the quenched galaxy number counts, as given by:
\begin{equation}
N^q(M_\star,z) = \sum_{\substack{i_{\rm gal}}} {\rm
  PDF}_{i_{\rm gal}}(M_\star, z) \times P^q_{i_{\rm gal}}(z).
\end{equation}

\noindent From there, we can compute the quenched number counts at the group redshift using equations~\ref{eq:Nbinned} and~\ref{eq:Ngroup}.

\subsection{Bayesian model}\label{bayesian_model}

The group galaxy quenched fraction is computed using Bayesian inference with the model presented in \citet{Andreon2006,DAgostini2004}. Using Bayesian inference allows us to properly account for the presence of two populations of galaxies ({\it field} $+$ {\it group}) with different quenched fractions in the region observed around each detected group. It is also free from the approximation that consists in confusing observed and true values of galaxy number counts, an approximation that may lead to negative number counts for the (unobserved) group population and quenched fractions outside the range $[0,1]$. The model we use can be written :
\begin{eqnarray}
\begin{aligned}
    {\rm obs}N_{\rm field} &\sim {\rm Poisson}({\rm true}N_{\rm field})\\
    {\rm obs}N_{\rm tot} &\sim {\rm Poisson}({\rm true}N_{\rm field} \times \frac{\Omega_{\rm group}}{\Omega_{\rm field}} + {\rm true}N_{\rm group})\\
    {\rm obs}N_{\rm field}^q &\sim {\rm Binomial}(f_{\rm field}^q,{\rm obs}N_{\rm field}) \\
    {\rm obs}N_{\rm tot}^q &\sim {\rm Binomial}(f_{\rm tot}^q,{\rm obs}N_{\rm tot})\\
    f_{\rm tot}^q &= \frac{f_{\rm field}^q\times{\rm true}N_{\rm field}\times \frac{\Omega_{\rm group}}{\Omega_{\rm field}} + f_{\rm group}^q\times{\rm true}N_{\rm group}}{{\rm true}N_{\rm field}\times \frac{\Omega_{\rm group}}{\Omega_{\rm field}} + {\rm true}N_{\rm group}},
\end{aligned}
\end{eqnarray}\label{eq:fQindiv}

\noindent where ${\rm obs}N_{\rm field},  {\rm obs}N_{\rm tot}$ are the observed galaxy number counts in the field reference region and group region respectively. These are modelled as being drawn from a Poisson distribution with parameters: ${\rm true}N_F,  {\rm true}N_{\rm tot}$, the true underlying galaxy number counts, in the field reference region and group region, respectively. The quantities ${\rm obs}N^q_{\rm field},  {\rm obs}N^q_{{\rm tot}}$ are the observed quenched galaxy number counts in the field reference region and group region respectively. These are modelled as being drawn from a Binomial distribution with probability of success (that is, the galaxy is quenched) $f^q_{\rm field}$,$f^q_{\rm tot}$ out of the ${\rm obs}N_{\rm field}, {\rm obs}N_{\rm tot}$ observed galaxy number counts, in the field reference region and group region respectively.

From these distributions, we then compute $f^q_{\rm group}$, the posterior for the quenched fraction of group galaxies considering we observed the sum of the two populations. We take uniform (flat) priors for all parameters (see online Appendix~{\textcolor{blue}{E} for details)}.
\noindent MCMC sampling of this model is performed using the {\small PYMC3} Python package.
\begin{figure*}
\centering
    \includegraphics[width=1.\textwidth]{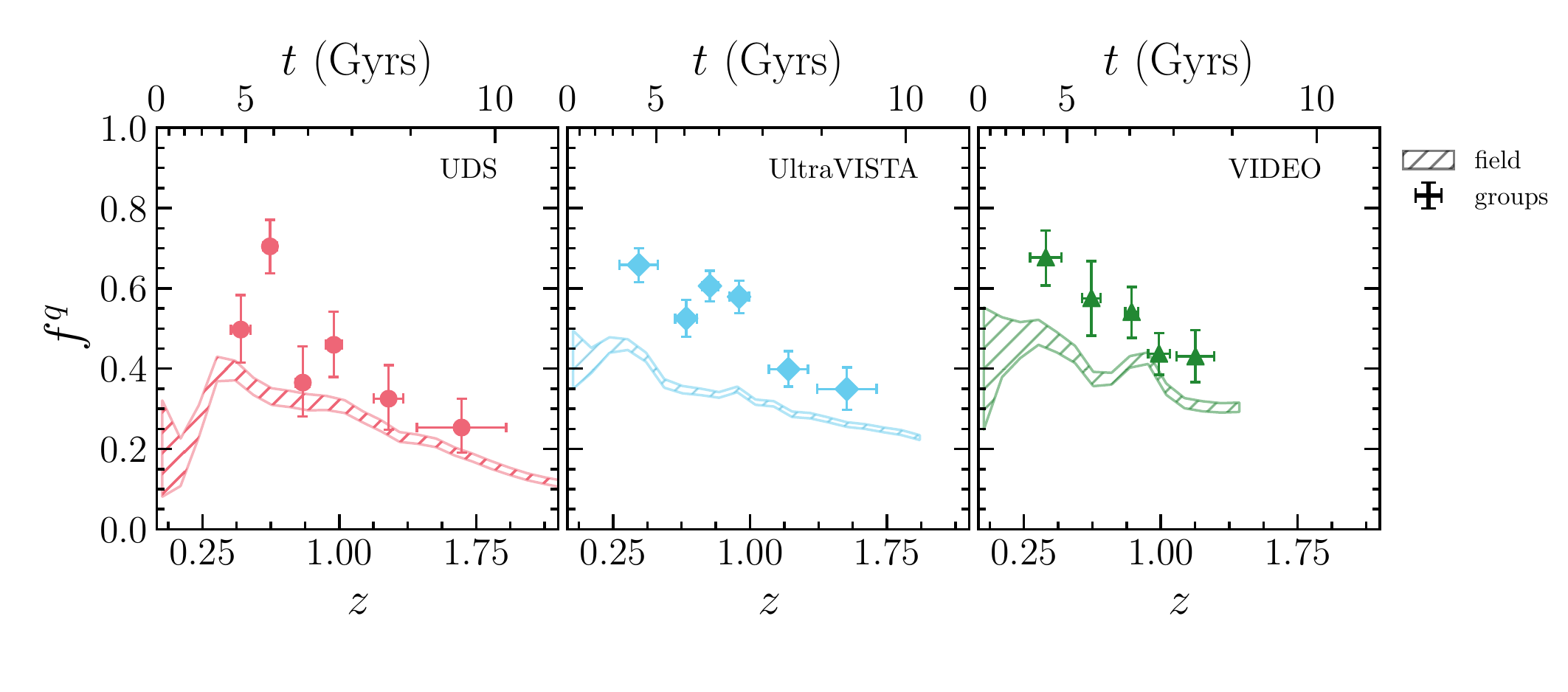}
    \vspace*{-1cm}
        \caption{Quenched fraction for galaxies with stellar-mass $10^{10.25} < M_\star / {\rm M}_{\sun} < 11$ in \detectifz~groups at $r < 0.5 \times R_{200}$. From left to right are shown the UDS (red dots), UltraVISTA (blue diamonds) and VIDEO (green triangles) survey regions. The point with error bar show the group quenched fraction in the redshift bins defined in Table~\ref{tab:fQbins}. Points are located at the median redshift of the bin and horizontal errorbars show the standard deviation of the redshift distributions of groups in the bin. Vertical errorbars are 68\% confidence limits in the $f_{\rm group}^q$ value. The dashed region shows the $68\%$ confidence region for the field quenched fraction $f^q_{\rm field}$.}
\label{fig:fQ_bins}
\end{figure*}

To compute the global field quenched fraction for each entire survey region (having removed regions around groups), we simply use the following model:
\begin{eqnarray}
\begin{aligned}
    {\rm obs}N_{\rm field} &\sim {\rm Poisson}({\rm true}N_{\rm field})\\
    {\rm obs}N_{\rm field}^q &\sim {\rm Binomial}(f_{\rm field}^q,{\rm obs}N_{\rm field}),
\end{aligned}
\end{eqnarray}
with flat priors for ${\rm true}N_{\rm field}$ and $f_{\rm field}^q$ (see online Appendix~\textcolor{blue}{E} for details).

\subsection{Fraction of quenched galaxies in individual groups}\label{sec:fq_indiv}

The fraction of quenched galaxies can then be computed for each individual galaxy group using the number counts and the Bayesian model presented in the two previous subsection. In practice, for each group, we need the total (group+field) counts and total quenched counts in the cluster region and reference counts from the fields. For the reference counts, as we want to use the group sample to infer properties of groups in general, we need to ensure that these counts are independent for each group, in order not to overestimate the strength of our results \citep[see][]{Raichoor2012}. To this aim, at each redshift step $dz=0.01$, we remove galaxies closer than $2 \times R_{200}$ from groups that may contribute to this redshift (whose ${\rm PDF}_{\rm group}(z) > 0$ at this redshift). Remaining galaxies are considered field galaxies at this redshift $z$. To keep independent counts for each group we use galaxies outside of our groups at each group's best redshift $z_{\rm group}$, and located in an annulus between $3$ Mpc and $5$ Mpc from the group centre. Using counts near the location of the group (local field) has the double advantage of ensuring independence of each group as well as accounting for correlated structures in the group vicinity.

\subsection{Stacking in redshift bins}\label{fQ:zbins}

To better understand the redshift evolution of the group quenched fraction $f_{\rm group}^q$, and compare the three survey regions, we stack individual groups in redshift bins. In our Bayesian inference framework, this comes to consider each individual group as one observation of the true underlying population in the redshift bin. 

For each of the three survey regions, we bin groups in 6 redshift bins. These bins were chosen to be roughly equally populated. The highest redshift bin is slightly different between UDS and UltraVISTA. This is due to the UltraVISTA stellar-mass completeness reaching $10^{10.25} {\rm M}_{\sun}$ at redshift $z=1.95$. For the same reason, there are no groups in the highest redshift bins for the VIDEO survey region. We have about 10, 40 and 15 groups per bin in the UDS, UltraVISTA and VIDEO regions respectively using this binning. The quenched fraction obtained are presented in Table~\ref{tab:fQbins} and Fig.~\ref{fig:fQ_bins}

\begin{table}
	\centering
	\caption{Quenched fraction in redshift bins}
	\label{tab:fQbins}
	\begin{tabular}{lccr} 
		\hline
		redshift range & $\langle z \rangle_{\rm median}$ & $f_{\rm group}^q$ & $N_{\rm group}$ \\
		\hline
		\hline
		UDS \\
		\hline
        $0.12 \le z < 0.52$ & 0.46 & $0.497^{+0.086}_{-0.080}$ & 11\\
		$0.52 \le z < 0.73$  & 0.62 & $0.705^{+0.066}_{-0.067}$  & 10\\
		$0.73 \le z < 0.88$  & 0.8 & $0.365^{+0.090}_{-0.085}$  & 11\\
        $0.88 \le z < 1.08$ & 0.97 & $0.460^{+0.082}_{-0.081}$ & 10\\
		$1.08 \le z < 1.42$ & 1.27 & $0.325^{+0.084}_{-0.078}$  & 9\\
		$1.42 \le z < 2.32$  & 1.66 & $0.253^{+0.072}_{-0.063}$  & 16\\
		\hline
		UltraVISTA \\
		\hline
        $0.12 \le z < 0.52$ & 0.39 & $0.658^{+0.041}_{-0.043}$ & 39\\
		$0.52 \le z < 0.73$  & 0.65 & $0.525^{+0.046}_{-0.046}$  & 41\\
		$0.73 \le z < 0.88$  & 0.78 & $0.606^{+0.038}_{-0.038}$  & 41\\
        $0.88 \le z < 1.08$ & 0.94 & $0.579^{+0.040}_{-0.041}$ & 40\\
		$1.08 \le z < 1.42$ & 1.21 & $0.399^{+0.045}_{-0.043}$  & 43\\
		$1.42 \le z < 1.95$  & 1.53 & $0.349^{+0.054}_{-0.052}$  & 38\\
        \hline
		VIDEO \\
		\hline
        $0.12 \le z < 0.52$ & 0.37 & $0.677^{+0.067}_{-0.070}$ & 16\\
		$0.52 \le z < 0.73$  & 0.62 & $0.575^{+0.093}_{-0.094}$  & 15\\
		$0.73 \le z < 0.88$  & 0.84 & $0.541^{+0.062}_{-0.064}$  & 15\\
        $0.88 \le z < 1.08$ & 0.99 & $0.437^{+0.052}_{-0.052}$ & 25\\
		$1.08 \le z < 1.42$ & 1.16 & $0.430^{+0.066}_{-0.064}$  & 20\\
	\end{tabular}
\end{table}

First, we note that the quenched fractions in the field and their redshift evolution are comparable in the three survey regions with an offset of $\sim 0.05$. We verified that the distribution in $\log {\rm SFR}$ vs $\log M_\star$ are similar for the three survey regions, as well as the normalisation and slope of the star-forming main sequence. We also verified that the offsets are not significant when accounting for cosmic variance using the {\small GETCV} IDL routine of \citet{Moster2011}.

In the entire redshift range probed here ($0.12 \le z < 2.32$) and each of the three surveys, the quenched fractions in the groups are found to be higher than in the field. However, due to the relatively small number of groups in each bin this trend is not very significant in some redshift bins in the UDS and VIDEO fields. Moreover, the group quenched fraction in redshift bins is similar between the different survey regions, with differences always smaller than $2.5 \sigma$ ($\sim 1\sigma$ in the mean). Considering the two previous points, we now study the group quenched fraction evolution jointly for the full REFINE survey.

\subsection{Quenched fraction redshift evolution in REFINE}\label{fQz}

To infer the redshift evolution of the quenched fraction, rather than fitting the binned data, we use a hierarchical Bayesian model. This has the advantage of avoiding a (somewhat) arbitrary choice of bins. This also allows us to account for uncertainty in the group redshift quite naturally. 
The adopted model is identical to the model presented in Sect.~\ref{bayesian_model} except the group quenched fraction now depends on the (unknown) true group redshift:

\begin{eqnarray}
\begin{aligned}
    f^q_{\rm group} &= {\tt ilogit} \left [ f_1 + \alpha_z (z^{\rm true}_{\rm group} - 1)\right],\\
    z^{\rm obs}_{\rm group} &= \mathcal{N}(\mu=z^{\rm true}_{\rm group}, \sigma=\sigma_{{\rm PDF}_{\rm group}(z)}^{68})
\end{aligned}
\end{eqnarray}\label{eq:fQz}

\noindent where ${\tt ilogit}(x) = (1 + e^{-x})^{-1}$ and ensures that $0 \le  f^q_{\rm group}  \le 1$ and $f_1$ and $\alpha_z$ are given flat priors. 
The observed group redshift is modelled as being drawn from a normal distribution, centred at the group true redshift, and with standard deviation equal to half the $68\%$ confidence interval around the median of the observed ${\rm PDF}_{\rm group}(z)$.

\begin{figure*}
\centering
    \includegraphics[width=0.9\textwidth]{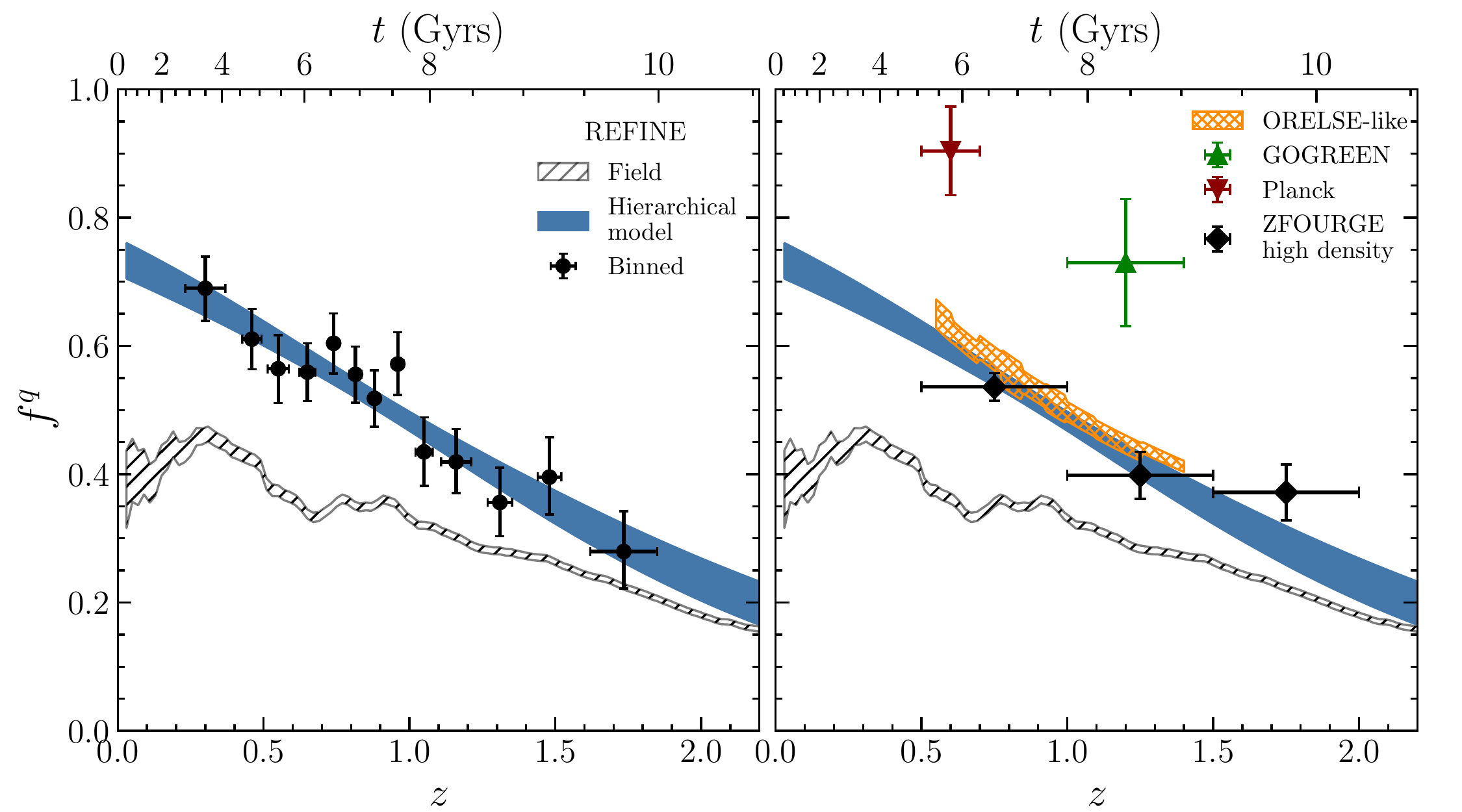}
        \caption{Quenched fraction for galaxies with stellar-mass $ 10^{10.25} < M_\star / {\rm M}_{\sun} < 11$ in \detectifz~groups at $r < 0.5 \times R_{200}$. Left: shows results for the full REFINE data. Points with errorbars show the binned group quenched fraction $f_{\rm group}^q$. Points are located at the median redshift of the bin, and the horizontal errorbars show the standard deviation of the redshift distributions of groups in the bin. Vertical errorbars are 68\% confidence limits in the $f_{\rm group}^q$ value. The shaded region displays the $68\%$ confidence interval on $f_{\rm group}^q(z)$ for the hierarchical model defined in Eq.~\ref{eq:fQz}. The dashed region shows the $68\%$ confidence region for the field quenched fraction $f^q_{\rm field}$. Right: $f^q_{\rm field}$ and $f_{\rm group}^q(z)$ from the left panel are reported and compared to values taken from the literature (ZFOURGE, ORELSE, Planck clusters and GOGREEN). See text for details.}
\label{fig:fQ_REFINE}
\end{figure*}

In addition to this hierarchical fit, we also computed the group quenched fraction in 13 equipopulated bins to allow for a sanity check of the goodness of fit. With this binning, we have $\sim 30$ groups per bins. Results are presented in the left panel of Fig.~\ref{fig:fQ_REFINE}. 
The group quenched fraction in our REFINE group sample is compatible with a linear decreasing redshift evolution, with $\alpha_z = -1.11^{+0.15}_{-0.16}$ and $f_1 = -0.07\pm 0.06$. The group quenched fraction of our REFINE group sample is found to be significantly higher than the field quenched fraction with higher confidence at low redshift. The confidence of the result monotonically drops with increasing redshift reaching $3\sigma$, $2\sigma$ and $1\sigma$ confidences at redshift $z = 1.55, 1.79$ and $2.23$ respectively. These results are discussed and compared to the literature in Sect.~\ref{sec:comparison_fQ_QFE}. We note that when running a similar hierarchical model (not presented here) for the global field quenched fraction on the REFINE sample, we also find it is compatible with a linear decreasing redshift evolution.

\subsection{Quenched fraction excess in REFINE}
\begin{figure}
\centering
    \includegraphics[width=0.42\textwidth]{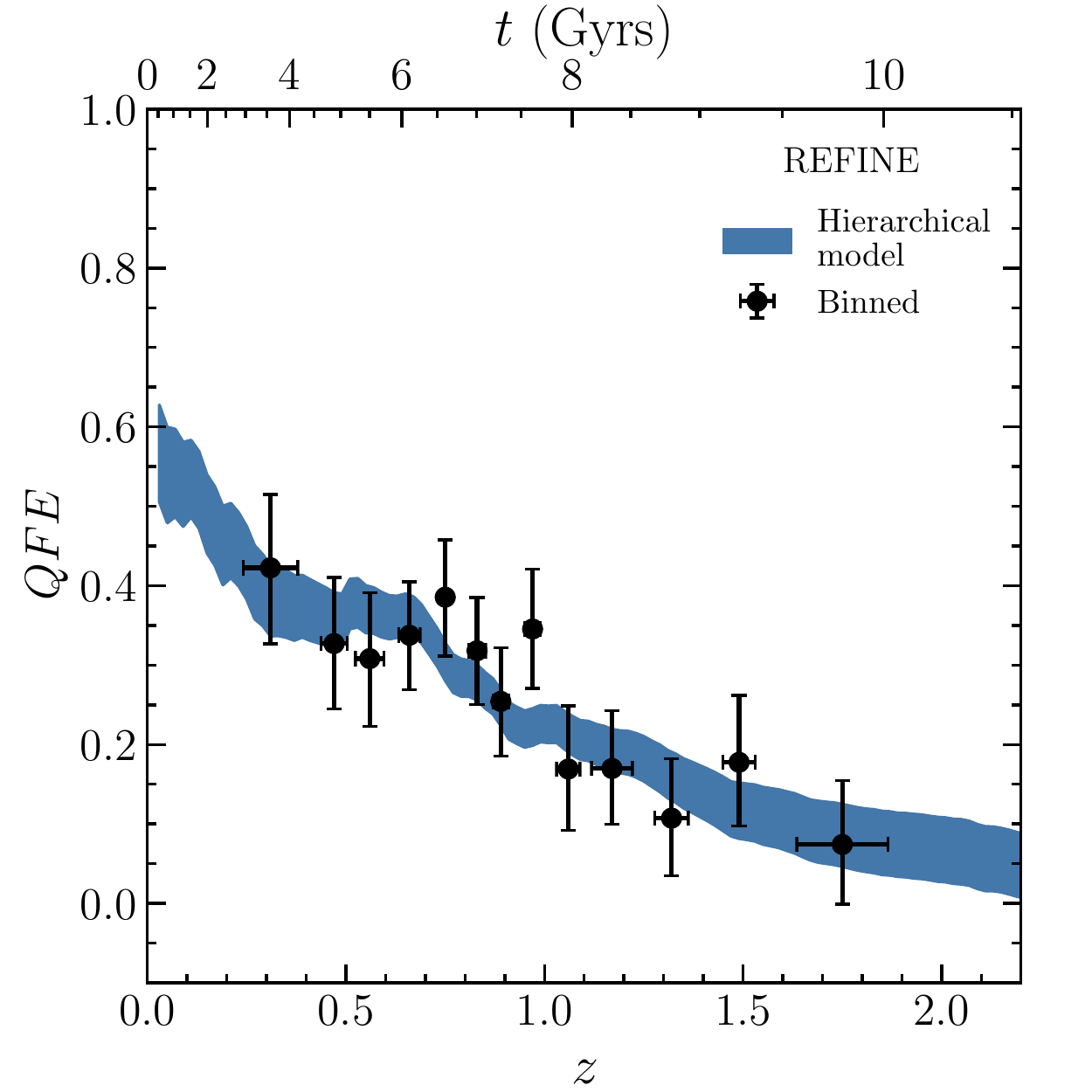}
        \caption{Quenched fraction excess ($QFE$) for galaxies with stellar-mass $ 10^{10.25} < M_\star / {\rm M}_{\sun} < 11$ in \detectifz~groups at $r < 0.5 \times R_{200}$. Points with errorbars shows the binned group quenched fraction ($QFE$). Points are located at the median redshift of the bin and horizontal errorbars show the standard deviation of the redshift distributions of groups in the bin. Vertical errorbars are 68\% confidence limits in the $QFE$ value. The shaded region shows the $68\%$ confidence interval on $QFE(z)$ obtained using $f^q_{\rm group}(z)$ from the hierarchical model defined in Eq.~\ref{eq:fQz}.}
\label{fig:QFE_REFINE}
\end{figure}

From the group and (global) field quenched fractions, we compute the quenched fraction excess ($QFE$) that describes the fraction of galaxies that would have been star-forming in the field but are quenched in their group environment following \citet{VdB2020}:
\begin{equation}
    QFE = \frac{f^q_{\rm group} - f^q_{\rm field}}{1 -  f^q_{\rm field}}.
\end{equation}
As noted in \citet{VdB2020}, the $QFE$ is also sometimes referred to as 'transition fraction' \citep[e.g.][]{vandenBosch2008}, 'conversion fraction' \citep[e.g.][]{Balogh2016} or 'environmental quenching efficiency' \citep[e.g.][]{Peng+10}.
In Fig.~\ref{fig:QFE_REFINE}, we present the $QFE$ obtained from our hierarchical fit of the group quenched fraction compared to the (global) field quenched fraction as a function of redshift, as well as its value in the 13 equipopulated bins presented in the previous section. We find that the $QFE$ decreases with increasing redshift, from $QFE = 0.53$ at $z=0.12$ to $QFE=0.03$ at $z = 2.31$, which respectively are the lowest and highest redshifts of our group sample. The effect of environment is significant (i.e $QFE > 0$) with $99.9\%$ and $95.4\%$ confidence up to $z = 1.59$ and $z=2.03$ respectively. These results are discussed and compared to the literature in Sect.~\ref{sec:comparison_fQ_QFE}.

\section{Discussion}\label{sec:discussion}

\subsection{Quenched and star-forming galaxies: sSFR cut}\label{sec:Discussion_sSFR}

We note that across the literature, many different cuts are used to select star-forming vs quenched galaxies (e.g. red-sequence: \citealp[][]{Popesso06}, SED fitting based classification: \citealp[][]{Sarron2018} or observed colour-colour: \citealp[][]{Bisigello2020}). A selection criterion that has gained in popularity is a rest-frame colour-colour cut based on the $UVJ$ diagram first proposed by \citet{WilliamsUVJ09} and then expended by other works using slightly different filters \citep[e.g. $NUVrK$:][]{ArnoutsNUVrK13}. 

In this work, we decided to use a cut in specific star-formation rate (sSFR) estimated from template fitting. We chose this selection criteria because in our probabilistic approach, it allows us to compute a probability that each galaxy is quenched at a given redshift (see Sect.~\ref{sec:Ncounts}). This way, our approach makes the quenched vs star-forming binary classification less strict and more resilient in degenerate cases.

We note that it has been showed that SED fitting SFR estimates are in general noisier and more biased than $NUVrK$ SFR estimates \citep[see][]{LeeN2015}. However, the discrepancy between both methods (bias) concerns mainly star-forming galaxies with ${\rm SFR} > 10^{1.5} {\rm M}_{\sun} {\rm yr}^{-1}$ which should not affect much the quenched vs star-forming binary classification. Moreover, with our probabilistic approach this possible mis-classification due to the higher scatter in SED fitting SFR estimates is mitigated.

In practice, we computed the probability for each galaxy to be quenched at each redshift, using a constant cut in sSFR (sSFR$< 10^{-11} {\rm yr}^{-1}$). \citet{Ilbert2013} showed that such a cut is similar to their colour-colour cut $NUV - r^+$ versus $r^+ - J$ at $z < 1$ and more conservative at $z > 1$. A similar trend was found in \citet{Ilbert2015} who found a sSFR$< 10^{-11} {\rm yr}^{-1}$ cut was very close to a colour-colour cut in $NUV - R$ vs $R-K$ out to $z = 1.4$.  We however investigate alternative constant cuts and find very similar results to what follows. Some authors \citep[e.g.][]{Lee2015,Jian2018} use instead a sSFR cut that evolves with redshift, to account for the general evolution of the galaxy star-forming main-sequence with redshift. However \citet{Lee2015} showed that this redshift dependent cut, even though it yields different quenched fraction values (higher at high redshift) does not affect the general trend of their results with galaxy stellar mass and environment. We investigated this and find largely the same result.

\subsection{Quenched fraction: Comparison to previous studies}\label{sec:comparison_fQ_QFE}

A number of studies computed the quenched fraction $f^q$ at different redshifts and stellar masses in different environments. Here we are interested in comparing our results with studies looking at similar redshifts and stellar-masses in high density environments, in particular at $z > 1$. We focus on the results presented in \citet{Papovich2018} at $0.15 < z < 2.0$ in the ZFOURGE survey \citep[][]{Straatman2016}, \citet{Lemaux2019} at $0.55 < z < 1.4$ in the ORELSE survey \citep[][]{Lubin2009}, \citet{VdB2018} at $0.5 < z < 0.7$ using Planck detected clusters and \citet{VdB2020} at $1.0 < z < 1.4$ in the GOGREEN survey \citep[][]{Balogh2017} as they provide either the galaxy stellar mass function (GSMF : ZFOURGE, Planck clusters and GOGREEN) or an analytical fit to the quenched fraction as a function of redshift, stellar mass and overdensity level (ORELSE), that allows for a fair comparison of our results to theirs (in particular using the same stellar-mass limits).

To probe the effect of environment on galaxy star-formation at high redshift, different approaches have been used in the literature. A common approach in deep contiguous surveys is to split the density field in four quartiles \citep[e.g. ZFOURGE][]{Papovich2018}. On the other hand one can instead observe some massive candidate galaxy clusters at different locations in the sky using dedicated observations \citep[e.g. GOGREEN][]{VdB2020}. The ORELSE survey took a hybrid approach, targeting specifically regions around massive clusters and studying regions with different overdensities in these fields \citep{Lemaux2019}.

Our own approach consists in detecting candidate groups in contiguous survey regions. Compared to a simple density cut, our method allows us to detect specifically structures that correspond to dark matter haloes, rather than galaxy filaments for example, that may be included in the highest density quartile of ZFOURGE. Owing to the relatively small volume covered by our three survey regions, our probability of observing massive clusters is low and we in fact do not observe very rich structures, contrary to the GOGREEN and ORELSE survey. Our study thus targets intermediate mass groups (at $0.12 \le z < 2.32$), a mass regime that is targeted by ORELSE (at $0.55 < z < 1.4$) but not by ZFOURGE, Planck and GOGREEN.

Comparison of our quenched fractions $f^q$ with those of the four works mentioned earlier are presented in the right panel of Fig.~\ref{fig:fQ_REFINE}. For ZFOURGE, Planck clusters and GOGREEN, quenched fractions are obtained using the empirical binned GSMF of the quenched and total galaxy population provided in each study. While the raw data is provided for Planck and GOGREEN, for ZFOURGE we extracted values of the data points from their figures using the WebPlotDigitizer tool \citep[][]{Rohatgi2020}. The quenched fraction $f^q$ is taken as the ratio of the quenched and total GSFM summed in the range $10.25 < \log M_\star / {\rm M}_{\sun} < 11$. For the ORELSE survey, we use the fit provided in \citet{Papovich2018} that gives $f^q$ as a function of redshift, stellar mass and overdensity $\log(1+\delta_{\rm gal})$. In each of the redshift bins used for the REFINE survey (showed in the left panel of Fig.~\ref{fig:fQ_REFINE}) that fall in the range $0.55 < z < 1.4$, we compute the median stellar mass ($\sim 10^{10.60} {\rm M}_{\sun}$) and $68\%$ interval of overdensity $\log (1+\delta_{M_\star})$ in $0.5 R_{200}$ using the \detectifz~density maps. Under the approximation $\delta_{M_\star} \simeq \delta_{\rm gal}$, this allow us to obtain an estimate of our sample quenched fraction using the ORELSE parametrization that we now refer to as the ORELSE-like quenched fraction. 

We find a very good agreement between our quenched fraction estimate and that of ZFOURGE and ORELSE-like. It should be noted that the ZFOURGE estimate at $1.5 < z < 2$ is $\sim 0.1$ higher than ours ($0.37 \pm 0.04$ vs $0.28 \pm 0.03$, $1.7 \sigma$ discrepancy). This discrepancy may come from different overdensities probed in this bin between ZFOURGE and our sample. We find that the ORELSE-like estimate are slightly higher than ours. The quenched fraction estimates however never differ by more than $1.3\sigma$ and around $0.5\sigma$ in the central part of the redshift ranged probed by the ORELSE survey. The slightly higher estimate for ORELSE-like parametrization could also be due to the fact that the quenched fraction estimates at $M_\star < 10^{10.5}$ (about half the weight of our sample) should be considered upper limits according to \citet{Papovich2018}. \\

\indent This relatively good agreement is particularly remarkable as ZFOURGE, ORELSE and our work used different cuts to segregate quenched and star-forming galaxies. This strengthens the argument made in Sect.~\ref{sec:Discussion_sSFR} that our probabilistic sSFR$< 10^{-11} {\rm yr}^{-1}$ cut is effectively equivalent to colour-colour cuts in $UVJ$ (ZFOURGE) and $NUVrJ$ (ORELSE) respectively.\\

On the other hand, there is a large discrepancy between our quenched fractions and those of \citet{VdB2018} and \citet{VdB2020} at redshift $0.5 < z < 0.7$ and $1.0 < z < 1.4$ respectively. Quenched fraction of the Planck clusters are 0.31 higher than ours ($0.90\pm0.07$ vs $0.59\pm0.01$, $4.4 \sigma$ discrepancy). Quenched fraction of the GOGREEN clusters are also $\sim 0.30$ higher than ours ($0.73\pm0.10$ vs $0.43\pm0.02$, a $3 \sigma$ discrepancy). 
To investigate the origin of this difference, for the GOGREEN survey, we use the cluster richness provided in \citet{VdB2020}. Using the \citet{VdB2020} definition of richness, our sample in $1.0 < z < 1.4$ has a median richness of 9.5, while GOGREEN clusters have a median richness of 35.8. For the Planck cluster sample, \citet{VdB2018} do not provide a richness estimate that we can compare to ours. However, their clusters have estimated masses $M_{500} > 5 \times 10^{14} {\rm M}_{\sun}$. Following the mass-richness scaling we found in our mock data, these massive clusters probe a very different halo mass range compared to our sample that is expected to be populated mostly with intermediate mass groups ($M_{200} \sim 5 \times 10^{13} {\rm M}_{\sun}$).

The galaxy quenched fraction is known to depend on cluster mass up to at least $z = 0.7$ \citep{Sarron2018}, with higher mass clusters having higher quenched fractions. While it is still unsure whether this result holds at higher redshift, it would explain our lower quenched fractions compared to \citet{VdB2018} and \citet{VdB2020}.

\subsection{Quenched fraction excess: redshift evolution}\label{sec:QFE_z}

We observe a decrease of the $QFE$ as a function of increasing redshift from $QFE = 0.53$ at $z = 0.12$ to $QFE = 0.035$ at at $z = 2.31$ (respectively the lowest and highest redshifts of groups used in this work). In order to interpret this result in terms of efficiency of quenching in galaxy groups, we need to ensure that we are comparing similar galaxy groups/overdensities at different redshifts and galaxies of similar stellar-masses. We note that, adopting this strategy, we are in fact looking at environments that are similar at different redshifts, thus not addressing the question of mass accretion on to haloes i.e. our high redshift groups are {\it not} the progenitors of our low redshift groups. The same is true for galaxies. We postpone such an analysis using the REFINE \detectifz~group sample to future work.

To check that we probe similar environments at all redshift, we verified that in the redshift bins showed in Fig.~\ref{fig:fQ_REFINE}, the groups have the same range of total stellar mass. We further checked that the typical $\delta_{M_\star}$ in $0.5 \times R_{200}$ is not redshift dependent. It is in fact constant around $\log (1 + \delta_{M_\star}) \sim 1 \pm 0.2$ (SD) in the range $0.12 < z < 2.32$.

To check that we study galaxies of similar stellar masses at different redshift, we verified that, in the redshift bins showed in Fig.~\ref{fig:fQ_REFINE}, galaxy stellar masses in the group and field sample have similar distributions and that these distributions are roughly independent of redshift. We find that the median stellar mass is 0.05 dex higher  in our groups than in the field overall ($10^{10.6} {\rm M}_{\sun}$ vs $10^{10.55} {\rm M}_{\sun}$), but the range covered by $68\%$ of the distributions are very similar $10.36 < \log M_\star/{\rm M}_{\sun} < 10.87$ vs $10.34 < \log M_\star/{\rm M}_{\sun} < 10.83$. The offset between median stellar masses in the group and field shows a slight redshift evolution from 0.04 at $z \sim 0.5$ to 0.09 dex at $z \sim 2$, but this shift is not significant considering the spread of the stellar mass distributions.

With the two previous checks, we can exclude that the observed increase of $QFE$ with decreasing redshift is due to observing more overdense groups at low redshift or galaxies of different stellar masses. However, as already mentioned, one needs to keep in mind that due to large-scale accretion (or equivalently halo mass growth) through cosmic time, galaxies in groups at $z \sim 2$ and $z \sim 0.5$ may have lived in different environments before entering the group. Indeed, the density contrast of the universe increases with cosmic time. This implies that galaxies in groups at lower redshift are expected to have spent more time in denser cosmic filaments than galaxies in groups at high redshift. Cosmic filaments have been shown to have higher quenched fraction than the field up to $z = 0.9$ \citep[see e.g.][]{Martinez2016,Salerno2019,Sarron2019} and to they play a specific role in quenching  \citep[see][]{Laigle2018,Kraljic2018}. Thus, without quantifying properly the level of pre-processing by cosmic filaments and its redshift evolution, we cannot conclude that the observed increase of $QFE$ with decreasing redshift is due to a specific quenching mechanism in the group environment whose efficiency might change with redshift. Computing the GSMF and radial distribution of quenched and star-forming galaxies should allow to address these points in more details \citep[see e.g.][]{VdB2018}.

\section{Conclusions}

In this paper we carry out a detailed investigation of groups and clusters in the three largest and deepest ground based surveys as part of the REFINE survey. Through a reanalysis of the redshifts and stellar masses of the UDS, COSMOS/UltraVISTA and CFHTLS-D1/VIDEO fields we develop a new methodology, called \detectifz~for finding clusters and groups up to $z \sim 3$ within these fields.   

To find these distant groups and clusters and determine cluster/group membership, our method uses the joint probability distribution functions for the stellar masses and redshifts of all the galaxies in the REFINE survey. We furthermore extensively test our methodology on simulation mocks which are designed to mimic the properties of the observational data that we use in this study.  Using these mocks, we show that we are able to retrieve a large fraction of all overdensities of galaxies and retrieve their properties accurately.

Overall, we detect and measure properties of 2588 candidate galaxy groups and clusters up to $z = 3.15$ at $S/N > 1.5$. We build a very pure ($> 90\%$) sample of 448 candidate groups up to $z=2.5$ (out of which 170 are newly detected) and study some of their properties, of which this paper is an initial investigation.

We use these results to investigate the quenched fraction of galaxies as a function of redshift and environment up to $z=2.31$.  We show that the differences between the quenched fraction in the field and in groups/clusters grows at lower redshifts, such that the differences in quenched fractions reach $\sim 0.3$ at $z \sim 0.1$, while it is $\sim 0.05$ at $z \sim 2$. As a result, we find that the quenched fraction excess ($QFE$) in groups grows at lower redshift from $QFE = 0.035$ at $z = 2.31$ to $0.53$ at $z=0.12$. 

We emphasize that these results are hard to interpret in terms of an increased quenching efficiency of the group environment at lower redshift at fixed halo mass/overdensity due to possible pre-processing happening outside the group environment (cosmic filaments) through cosmic time. 
We plan to explore this aspect in more details through an analysis of the group galaxy stellar-mass function and radial profile up to several virial radii in future work.

\section*{Acknowledgements}

We thank Ulrike Kuchner for useful discussions, and thank the various teams that produced the initial images we use from UltraVISTA, UDS and VIDEO.  This study was support by a Cosmic Visions STFC grant and by the Universities of Manchester and Nottingham.

\section*{Data Availability}

The \detectifz~Python package used to detect galaxy groups will be made available at https://github.com/fsarron/detectifz, along with the group and group member catalogues. The group catalogues will also be released at the CDS. The remaining data underlying this article, including galaxy catalogues, mock galaxy catalogues, galaxy physical parameter estimate PDF and their by-products (e.g. galaxy number counts, quenched fractions, etc.) will be shared on reasonable request to the corresponding author.




\bibliographystyle{mnras}
\bibliography{sarron_biblio} 



\section*{Supporting Information}\label{Supp}
Supplementary data are available at MNRAS online.\\

\noindent \textbf{Appendix~D} Group mass proxy : comparison to literature\\
\noindent \textbf{Appendix~E} Quenched fractions : Priors of Bayesian models\\

\noindent Please note: Oxford University Press is not responsible for the
content or functionality of any supporting materials supplied by
the authors. Any queries (other than missing material) should be
directed to the corresponding author for the article.

\appendix
\section{Examples of \detectifz~groups in the UDS field}\label{sec:appendix:iJK}

\begin{figure*}
    \includegraphics[width=1.05\textwidth]{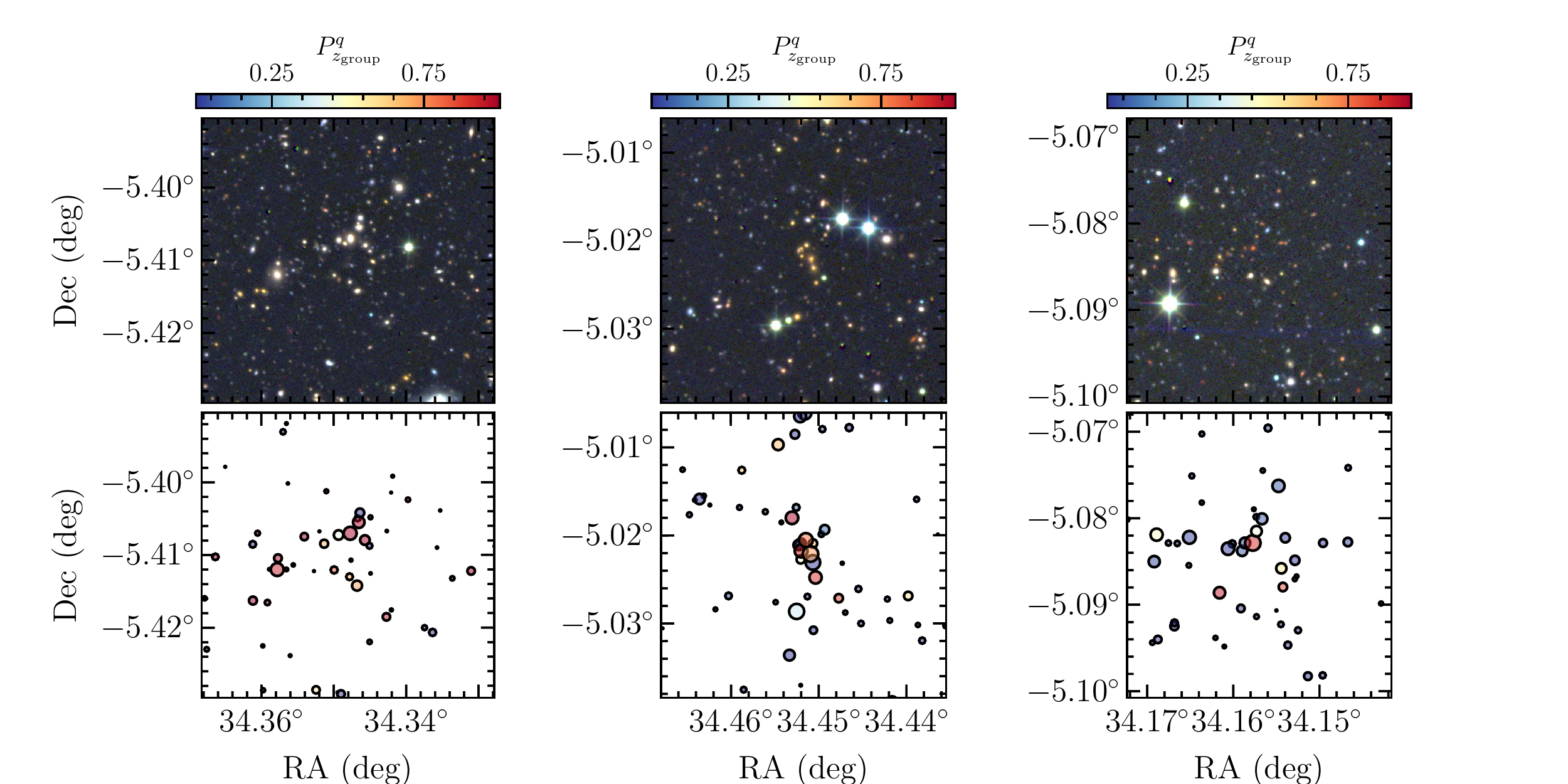}
        \caption{Top row : From left to right, $iJK$ images of three clusters located at $z=0.6, 1.28$ and $2.20$, respectively. Bottom row :  galaxies in the group field of view at a distance $d < R_{200,M_\star}$ from the group centre and with $P_{\rm mem} > 0.2$ colour-coded by their probability to be quenched at the group redshift. The point size is proportional to the galaxy stellar mass at the group redshift. See text for details.}
\label{fig:iJK}
\end{figure*}

In Fig.~\ref{fig:iJK} we show example $iJK$ images and some properties of galaxies near the identified group centre. The three groups were detected in the UDS survey region at redshifts $z=0.6, 1.28$ and $2.20$ respectively, at signal-to-noise ratio $S/N \ge 5$. The $iJK$ images are centred on each \detectifz~group's centre. The bottom row shows galaxies in the group field of view at a distance $d < R_{200,M_\star}$ from the group centre and with $P_{\rm mem} > 0.2$. The point size is proportional to the galaxy median stellar mass at the group redshift $M_\star^{{\rm median},z_{\rm group}} = \langle \int {\rm PDF}_{\rm gal}(M_\star,z)~{\rm PDF}_{\rm group}(z)~dz\rangle_{\rm median}$ and colour-coded with the probability that the galaxy is quenched if located at the group redshift $P^q_{z_{\rm group}} = \int P^q(z)~{\rm PDF}_{\rm group}(z)~dz$, similarly to the quantities used for delocalised number counts in Sect.~\ref{sec:Ncounts}. We note that the group at $z=1.28$ is newly detected by \detectifz.

\section{Technical details on mocks}\label{sec:apendix:mock}

In this appendix, we give the technical details of the method we used to build mock data resembling REFINE data from the lightcones of \citet{Henriques15}, that were briefly presented in Sect.~\ref{data:mock}.

\subsubsection{Survey geometry and $K$ band magnitude}

We started by shifting the lightcones sky coordinates to match those of
the survey we want to mimic and applied the masks due to bright stars
in the field of the surveys to
reproduce each survey geometry. We then proceed to add photometric-like noise to the true values of
$K$ band magnitude $m_K$, redshift $z$ and stellar-mass $M_\star$. For the magnitude, we computed the median $1\sigma$ uncertainties
$\langle \sigma_{m_{K}\rangle_{\rm median}}(m_K,z)$ on $m_K$ in
bins of $m_K$ and $z$ in the data. We binned the lightcone galaxies in
the same bins and added a shift to the true magnitude sampled from a
normal distribution such that
\begin{equation}
{\rm obs}m_K \sim \mathcal{N}(\mu=m_{K,{\rm
    true}},~\sigma=\langle \sigma_{m_{K}\rangle_{\rm median}}(m_K,z)).
\end{equation}
\noindent We keep in the mock only galaxies with ${\rm obs}m_K < m_K^{90\%}$.  This ensures that we are dealing with a complete sample and that fainter galaxies than our observational limit are not included in the computation.  This process is therefore sensitive to the survey region considered, and as such we have to create these mocks for each of our REFINE fields, mimicking the observational data in each one.

\subsubsection{The ${\rm PDF}(M_\star,z)$ and $(M_\star,z)$ shift}

We want each mock galaxy to have its redshift and
stellar-mass shifted from its true value with a shift typical of what
is expected for such a galaxy in the data. Each mock galaxy should
also have a ${\rm PDF}(M_\star,z)$ typical of what is expected for
such a galaxy in the data.

To do this, we binned the data in 2D bins $(M_\star^{\rm median},
z_{\rm phot})$ with stellar-mass and redshift steps of $dM_\star =
0.05$ and $dz = 0.01$ as for ${\rm PDF}(M_\star,z)$. To avoid too
sparse sampling and because of the typical uncertainty on these points
estimates of $M_\star$ and $z$, the size of the bins are taken to be $\Delta
M_\star = \langle \sigma^{68}_{M_\star}\rangle_{\rm median} = 0.2$ and $\Delta z = \sigma_0 \times (1+z)$, with $\sigma_0 = 0.045, 0.01, 0.035$ for UDS, UltraVISTA and VIDEO respectively, effectively applying a running window with overlapping bins. Similarly, we binned mock galaxies in 2D bins $(M_\star^{\rm true},
z^{\rm true})$  with stellar-mass and redshift steps of $dM_\star =
\Delta M_\star = 0.05$ and $dz = \Delta_z = 0.01$. Here the bins are not
overlapping as $M_\star^{\rm true}$ and $z^{\rm true}$ are known exactly.

Each bin $(M_\star,z)_{\rm bin}$ is populated with $N_{\rm gal,bin}^{\rm
  data}$ and $N_{\rm gal,bin}^{\rm mock}$. We bootstrap $N_{\rm gal,bin}^{\rm
  mock}$ from the list of $N_{\rm gal,bin}^{\rm
  data}$ data galaxies. This effectively gives us a
${\rm PDF}(M_\star,z)$ for each of these mock galaxies. We note that some ($< 1\%$) mock
galaxies have $(M_\star^{\rm true},z^{\rm true})$ values that do not
appear in the data. We assign to these 2D normal PDFs centred at the
bin values, with standard
deviations $(\langle \sigma^{68}_{M_\star}\rangle_{\rm median},\sigma_0 \times (1+z^{\rm
  true}))$ and no covariance.

Each mock galaxy ${\rm PDF}(M_\star,z)$ is re-centred at the central value
of the bin $(M_\star,z)_{\rm bin}$. We use the {\small PINKY} Python
package to sample a given $(\widehat{M}_\star^{\rm
  median},\widehat{z}_{\rm phot})$ from the PDF, and re-centre the PDF
at these sampled values.

This last ${\rm PDF}(M_\star,z)$ is our final
estimate for the mock galaxies. It is sampled directly from the data
and has been shifted to account for the uncertainty information it
encompasses. \\

\section{Technical details on the probabilistic membership assignment}\label{sec:appendix:pmem}

\subsection{Bayesian formalism}\label{detectifz:pmem:bayes}
As mentioned in Sect.~\ref{DETECTIFz:pmem}, the probabilistic membership of group members is computed in a Bayesian formalism
similar to that developed in \citet{George11} and \citet{CBprobamem}
in which the probability for a galaxy (${\rm gal}$) to be a member of
a given group ($G$) is defined as the posterior probability $P_{\rm mem}
\equiv P({\rm gal} \in G \vert {\rm PDF}_{\rm gal}(M_\star,z),~{\rm PDF}_{\rm group}(z))$. In all generality, we should adopt a cluster model specifying
the group galaxy stellar-mass function (GSMF) and redshift distribution to which the
observed ${\rm PDF}_{\rm gal}(M_\star,z)$ is to be compared. Keeping in
line with the \detectifz~idea of being model free, as models (in particular group GSMF) are not yet well constrained at low group mass and high redshift, we instead reduce constraints on $M_\star$ by marginalising it out, and consider only the redshift information ${\rm PDF}(z)$. The redshift distribution is modelled by the $\delta$ distribution in redshift peaked at $z_{\rm group}$. As the photometric redshift uncertainty is an order of magnitude larger than the actual group size in redshift space, this makes a more sophisticated model unnecessary for our purpose.

So in practice we can write $P_{\rm mem}$ using Bayes theorem and focusing on the redshift information:
\begin{eqnarray}
  \begin{aligned}
    P_{\rm mem} \propto \int P({\rm PDF}_{\rm gal}(z) \vert {\rm gal} \in
G,~{\rm PDF}_{\rm group}(z),~M_\star^{\rm gal})(z)\\
    \times ~P({\rm gal} \in G~\vert~{\rm PDF}_{\rm group}(z),~M_\star^{\rm gal})(z)~dz,
\end{aligned}
\end{eqnarray}
\noindent where $P({\rm PDF}_{\rm gal}(z) \vert {\rm gal} \in G,~{\rm PDF}_{\rm group}(z),~M_\star^{\rm gal})(z)$ is the likelihood of observing the probability ${\rm PDF}_{\rm gal}(z)$ at redshift $z$ knowing that it is a member of the group $G$ and $P({\rm gal} \in G~\vert~{\rm PDF}_{\rm group}(z),~M_\star^{\rm gal})(z)$ is the prior probability that the galaxy belongs to the group evaluated at a redshift $z$. Note that these are expressed as PDFs of redshift, and the probability that the galaxy belongs to the group (up to some constant, see Sect.~\ref{detectifz:pmem:norm}) is obtained by integrating their product over all $z$.

\subsection{Likelihood}\label{detectifz:pmem:likelihood}

The likelihood can be
computed through a convolution of the observed ${\rm PDF}_{\rm
  gal}(z)$ with an expected  ${\rm PDF}_{\rm
  gal \in G}(z)$ for a group galaxy. Given our model and the photometric redshift uncertainty $\sigma_{z}^{68}(M_\star,z)$, the probability of observing a true group galaxy of stellar mass $M_\star^{\rm gal}$ located at the redshift of the group ($z_{\rm group}$) can be approximated by a 1D normal distribution in the redshift dimension :
\begin{eqnarray}
  \begin{aligned}
   \mathcal{N}_G(z~\vert~M_\star^{\rm gal}) = \mathcal{N}[z,
   \mu&=z_{\rm group},\\
    \sigma&=\langle\sigma_{z}^{68}\rangle_{P_{68}}(M_\star^{\rm gal},z_{\rm group})].
     \end{aligned}  
\end{eqnarray}\label{eq:like_normal}

\noindent In addition to the photometric uncertainty, we also have an
uncertainty on the cluster redshift expressed through ${\rm PDF}_{\rm group}(z)$. such that the probability of observing a true group galaxy of stellar mass $M_\star^{\rm gal}$ at a given redshift $z$ is obtained through a convolution of the two distributions:
\begin{eqnarray}
  \begin{aligned}
    {\rm PDF}_{\rm{gal} \ \in \ G}(z~\vert~M_\star^{\rm gal}) = &~({\rm PDF}_{\rm group}(z) *
    \mathcal{N}_G(z~\vert~M_\star^{\rm gal})\\ = &\int {\rm
      PDF}_{\rm group}(z-z') \mathcal{N}_G(z~\vert~M_\star^{\rm gal}) \ dz'.
  \end{aligned}  
\end{eqnarray}

\noindent The likelihood of observing a given ${\rm PDF}_{\rm gal}(z)$ if a galaxy of mass $M_\star^{\rm gal}$ is a group member then is:
\begin{eqnarray}
  \begin{aligned}
    P({\rm PDF}_{\rm
  gal}&(z)~\vert~{\rm gal} \in G,~{\rm PDF}_{\rm group}(z))(z) =\\ &
{\rm PDF}_{\rm gal}(z) \ {\rm PDF}_{\rm{gal} \ \in \
  G}(z \vert M_\star^{\rm gal})  ,
\end{aligned}
\end{eqnarray}
\noindent where $M_\star^{\rm gal}$ is taken as the median of $\int {\rm PDF}_{\rm gal}(M_\star,z)~{\rm PDF}_{\rm group}(z)~dz$ (median stellar mass at the group redshift).

\subsection{Prior}\label{detectifz:pmem:prior}

\noindent The prior probability $P^{\rm prior}(z\vert M_\star^{\rm gal}) \equiv P({\rm gal} \in G~\vert~{\rm PDF}_{\rm group}(z), M_\star)(z)$ is obtained by quantifying the number count excess at the galaxy position for each $z$. To this aim we first need to compute galaxy number counts at the group redshift $z_{\rm group}$ knowing the uncertainty on this value encoded in ${\rm PDF}_{\rm group}(z)$.

The number count excess at redshift $z$ for a galaxy of stellar mass $M_\star^{\rm gal}$ is taken as : 
\begin{equation}
  P^{\rm prior}(z\vert M_\star^{\rm gal}) =  1 - \frac{f_{\rm loc} \times N_{F_{\rm noclus}}(M_\star^{\rm gal},z) \times \frac{\Omega_{R_{200}}}{\Omega_{F_{\rm noclus}}(z_{\rm group})}}{f_{\rm sky} \times N_{{\rm tot},R_{200}}(M_\star^{\rm gal},z)},
\end{equation}

\noindent where number counts $N(M_\star,z)$ are computed as outlined in Sect.~\ref{data:SMpy}, with $\left [ M_{\rm inf}, M_{\rm sup}\right ] = [M_\star^{l95}, M_\star^{u95}]$ and $\left [ z_{\rm inf}, z_{\rm sup}\right ] = z_{\rm group} \pm \sigma_{z}^{95}$ \citep[see also][for a similar running window]{CBprobamem}. $N_{F_{\rm noclus}}(M_\star,z)$ are the {\it field} number counts, computed in $\Omega_{F_{\rm noclus}}(z)$ area i.e. keeping only galaxies that are at a distance $d > 2 \times R_{200}$ from \detectifz~detected groups that may contribute at the group redshift $z_{\rm group}$. $N_{{\rm tot},R_{200}}$ are the {\it total (group + field)} number counts at $d < R_{200}$ from the group centre.

There are two additional correction factors in this equation : $f_{\rm loc}$ and $f_{\rm sky}$. The former allows us to correct for the local large-scale structure bias around the cluster that may be higher or lower than the mean field value obtained from the full survey ($F_{\rm noclus}$), similarly to \citet{CBprobamem}. The latter allows us to account for the cluster profile, giving higher $P^{\rm prior}$ values to galaxies located in the denser regions of the group. Both are computed using the density maps generated in Sect.~\ref{DETECTIFz:DTFEMC} at the group best redshift $z_{\rm group}$. The value $f_{\rm loc}$ is the ratio between the mean pixel value in an annulus centred at the group centre with inner radius $3$ Mpc and outer radius $5$ Mpc, and the mean pixel value over the all field (with the clusters removed). Furthermore,
$f_{\rm sky}$ is the ratio between the value of the pixel in which the galaxy is located and the mean pixel value in $ R_{200}$.

\subsection{Normalisation and final probability of membership}\label{detectifz:pmem:norm}

To form our final $P_{\rm mem}$ estimates, we need to normalise the numerator of the right-hand side of Bayes theorem equation. In the formalism presented here, different schemes of normalisation have been proposed in the literature. In particular, \citet{George11} consider that each galaxy  is either a group member of a field member and thus use a Bayesian formalism with two events $A \equiv {\rm gal} \in G$ or $ B \equiv {\rm gal} \in F$. They thus normalise their estimates such that the total probability of the event $A \vert \vert B$ is equal to one.
\citet{CBprobamem} instead proposed to normalise the probability of each galaxy to the maximum probability it may reach i.e. the probability a fiducial group member would have in the case of a negligible field. \citet{CBprobamem} showed that their re-scaling is more efficient than the \citet{George11} estimator. Thus, we choose to adopt the \citet{CBprobamem} method in this work. In practice this is done for each galaxy by shifting its ${\rm PDF}_{\rm gal}(z)$ so that it peaks at the same redshift as ${\rm PDF}_{\rm group}(z)$ and fixing $\sigma_z = 0.01$ in Eq.~\ref{eq:like_normal}:
\begin{equation}
    P_{\rm mem} = \frac{ \int
{\rm PDF}_{\rm gal}(z) \ {\rm PDF}_{\rm{gal} \ \in \
  G}(z \vert M_\star^{\rm gal})  P^{\rm prior}(z\vert M_\star^{\rm gal}) dz}{\int
{\rm PDF}_{\rm gal}(z-z_{\rm group}) \ {\rm PDF}_{\rm{gal} \ \in \
  G}(z\vert\sigma_z = 0.01)dz},
\end{equation}

\section{Group mass proxy : comparison to literature}\label{sec:apendix:mass}

\begin{figure*}
\centering
\includegraphics[width=1.0\textwidth]{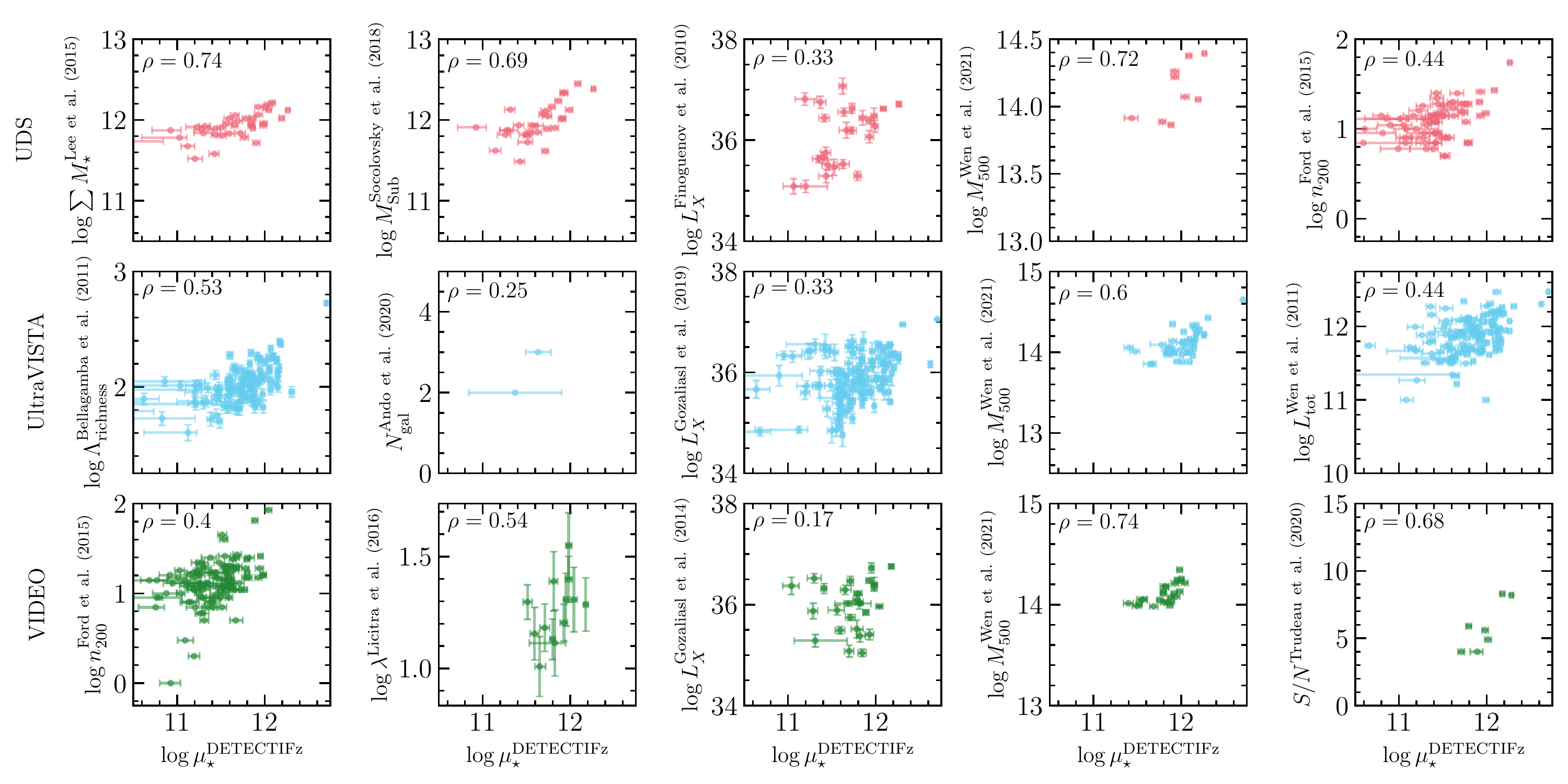}
    \caption{Group mass proxies from the literature vs $\log \mu_\star^{\rm \detectifz}$ for groups in common between our catalogue and catalogues from the literature. The top, middle, and bottom rows show groups matched in the UDS, UltraVISTA and VIDEO respectively. See Sect.~\ref{sec:match} for detail on the matching process. Among the matched catalogues that provide a mass proxy, we only include the five with the most groups in common with \detectifz. The Spearman correlation coefficient $\rho$ is indicated in each panel.}
    \label{fig:mass_proxies}
\end{figure*}


\noindent In Fig.~\ref{fig:mass_proxies} we compare a series of properties, including our group mass proxy - the group total stellar mass $\log \mu_\star$, to mass proxies in group and cluster catalogues from the literature that we matched to ours. Catalogue matching is performed using two-way geometrical matching (see Sect.~\ref{sec:match} of the paper for details) with a maximum offset of 1 Mpc on the sky (at the redshift of the  \detectifz~group) and $\vert \Delta z \vert = 2\times \langle\sigma_z^{68}\rangle_{P_{68}}(z)$ in redshift space. The mass proxies used within the studies we compare our catalogue to have very different definitions (e.g. mass estimated from richness, X-ray luminosity, etc). They are thus expected to have different correlations with the true total group mass, with a varying amount of scatter. This makes studying closely the relation between the different proxies beyond the scope of this work. However, we plot the literature proxies against $\log \mu_\star^{\rm \detectifz}$ for the three survey regions in Fig.~\ref{fig:mass_proxies} for the interested reader.

In each survey region, among the matched catalogues that provide a mass proxy, we plot only the five with the most groups in common with \detectifz. In each panel the Spearman correlation coefficient is indicated. In the case of X-ray matched catalogues, we only plot the comparison of the group mass proxies for \detectifz~groups whose centre is located inside the X-ray $R_{200}$, as these are the most secure matches. When looking at the Spearman correlation coefficients, we see that \detectifz~mass proxy $\log \mu_\star$ has a positive correlation with all other mass proxies from the literature, with values going from $\rho = 0.17$ (weak correlation) to $\rho = 0.74$ (strong correlation). We note that the Spearman correlation coefficient does not account for uncertainties. However, as can be seen our values compare very well with similar quantities measured in different ways from previous work.

\vspace*{2cm}

\section{Quenched fractions : Priors of Bayesian model}\label{sec:appendix:BayesPriors}
We present below the priors used for each variable in the models presented in Sect.~\ref{bayesian_model} and Sect.~\ref{fQz} of the paper.\\

\noindent For the model presented in Eq.~ \ref{eq:fQindiv} of the paper, we use the following priors:
\begin{eqnarray}
\begin{aligned}
    {\rm true}N_{\rm field} &\sim {\rm Uniform}(1,\infty)\\
    {\rm true}N_{\rm group} &\sim {\rm Uniform}(1,\infty)\\
    f_{\rm field}^q &\sim  {\rm Uniform}(0,1)\\
    f_{\rm group}^q &\sim  {\rm Uniform}(0,1).
\end{aligned}
\end{eqnarray}

\noindent For the model presented in Eq.~ \ref{eq:fQindiv} of the paper, we use the following priors:
\begin{eqnarray}
\begin{aligned}
    {\rm true}N_{\rm field} &\sim {\rm Uniform}(1,\infty)\\
    f_{\rm field}^q &\sim  {\rm Uniform}(0,1).
\end{aligned}
\end{eqnarray}

\noindent For the hierarchical model presented in Eq.~ \ref{eq:fQz} of the paper, we use the following priors:
\begin{eqnarray}
\begin{aligned}
    {\rm true}N_{\rm field} &\sim {\rm Uniform}(1,\infty)\\
    {\rm true}N_{\rm group} &\sim {\rm Uniform}(1,\infty)\\
    f_{\rm field}^q &\sim  {\rm Uniform}(0,1)\\
    z_{\rm group}^{\rm true} &\sim  {\rm Uniform}(0,5)\\
    f_1 &\sim  \mathcal{N}(\mu=0,\sigma=10)\\
    \alpha_z &\sim  \mathcal{N}(\mu=0,\sigma=10).
\end{aligned}
\end{eqnarray}

\bsp	
\label{lastpage}
\end{document}